\documentclass{aa}  

\usepackage{graphicx}
\usepackage{txfonts}
\usepackage{hyperref}
\hypersetup{
    colorlinks = true,
    linkcolor = blue,
    filecolor = blue,      
    urlcolor = blue,
    citecolor = blue,
}
\usepackage{multirow}

\begin{document}

   \title{Non-modulated pyramid wavefront sensor}
   \titlerunning{Non-modulated pyramid WFS}

   \subtitle{Use in sensing and correcting atmospheric turbulence}


   \author{G. Agapito
          \inst{1}
          \and
          E. Pinna\inst{1}
          \and
          S. Esposito\inst{1}
          \and
          C. T. Heritier\inst{2,3}
          \and
          S. Oberti\inst{4}
          }

   \institute{Osservatorio Astrofisico di Arcetri (INAF), Largo E. Fermi 5, Firenze, Italy, 50125\\
              \email{guido.agapito@inaf.it}
         \and
             DOTA, ONERA [F-13661 Salon cedex Air - France] 
         \and
            Aix Marseille University, CNRS, CNES, LAM, Marseille, France
         \and
            European Southern Observatory, Karl-Schwarzschild-str-2, D-85748 Garching, Germany
             }

   \date{Received XXX; accepted YYY}

 
  \abstract
   {The diffusion of adaptive optics systems in astronomical instrumentation for large ground-based telescopes is rapidly increasing and the pyramid wavefront sensor is replacing the Shack-Hartmann as the standard solution for single conjugate adaptive optics systems. The pyramid wavefront sensor is typically used with a tip-tilt modulation to increase the linearity range of the sensor, but the non-modulated case is interesting because it maximizes the sensor sensitivity. The latter case is generally avoided for the reduced linearity range that prevents robust operation in the presence of atmospheric turbulence.}
   {We aim to solve part of the issues of the non-modulated pyramid wavefront sensor by reducing the model error in the interaction matrix. We linearize the sensor response in the working conditions without extending the sensor linearity range.}
   {We developed a new calibration approach to model the response of pyramid wave front sensor in partial correction, whereby the working conditions in the presence of residual turbulence are considered.}
   {We use in simulations to show how the new calibration approach allows for the pyramid wave front sensor without modulation to be used to sense and correct atmospheric turbulence and we discuss when this case is preferable over the modulated case.}
   {}



   \maketitle
%

\section{Introduction}\label{intro}

\textcolor{black}{Adaptive optics (AO) for astronomy is a well-developed technique that is meant to compensate for the effects of atmospheric distortion. Currently, several instruments in the 8m-class telescope are assisted by this technique: i.e., SPHERE \cite{2019A&A...631A.155B}, MUSE+GALACSI \cite{2020SPIE11448E..0VH}, ERIS \cite{2023arXiv230402343D}, OSIRIS \cite{2006SPIE.6269E..1AL}, KAPA \cite{2022SPIE12185E..0QW}, SCEXAO \cite{2020SPIE11448E..7HC}, LUCI \cite{2018SPIE10702E..0BH}, however, this is not an exhaustive list.
The next generation of optical and infrared instruments for ground-based telescopes will also be equipped with AO modules to improve the image quality of the scientific instruments \citep[i.e.,][again, this is not exhaustive list]{2021Msngr.182...17D,2021Msngr.182...13C,2021Msngr.182....7T,2021Msngr.182...33H,2021Msngr.182...22B,2021Msngr.182...27M,2018SPIE10703E..3VC,2018SPIE10703E..0WB}. While the current generation of AO systems generally includes a Shack-Hartmann wavefront sensor \citep[SHS,][]{hartmann}, most of the future single conjugate AO (SCAO) systems are designed with a pyramid wavefront sensor \citep[PWFS,][]{Ragazzoni96}.
This sensor exhibits a better sensitivity than the SHS as shown in \cite{2004OptCo.233...27V} and proved on sky to produce a high level of wavefront correction on a large range of natural guide star (NGS) magnitudes \citep[see][]{Esposito2010,Pinna2019}.}

One of the key characteristics of the PWFS is the ability to tune the sensitivity and the linearity range through tip-tilt modulation.
In current SCAO systems, the PWFS is used with a tip-tilt modulation radius of a few $\lambda/D$ (where $\lambda$ is the sensing wavelength and $D$ is the telescope diameter) to increase the linearity range of the sensor at the cost of a reduced sensitivity.
In \cite{2001A&A...369L...9E}, it was shown that the wavefront reconstruction error variance for a non-modulated PWFS
 is higher than the one of a modulated PWFS, except in cases of low flux and small input variance levels.
These and similar results \citep[see e.g.,][]{2022A&A...658A..49B} have persuaded the designers of AO PWFS-based systems to discard the non-modulated configuration.
\textcolor{black}{However}, the non-modulated PWFS, as \textcolor{black}{demonstrated} in \cite{2004OptCo.233...27V}, is able (in principle) and excluding \textcolor{black}{non-linearity}, to correct almost perfectly the \textcolor{black}{lower} spatial frequencies and to offer an increased sensitivity that is particularly \textcolor{black}{attractive} for extreme AO (XAO) systems and for differential piston sensing.
In fact, for example, \cite{2022JATIS...8a9001C} proposed a second stage AO system with a non-modulated PWFS, \cite{2014SPIE.9148E..2MP} proposed a PWFS-based system with a small modulation radius to balance atmospheric and differential segment piston modes correction, while \cite{2022JATIS...8b1502E} proposed a mixed approach where PWFS modulation alternates between modulated and non-modulated.
In the latter case, the measurement done during non-modulated frames are used for the correction of differential segment piston modes.

The limitations of non-modulated PWFS shown in \cite{2001A&A...369L...9E} are real, but in this paper we show that a different calibration of such a sensor can help to overcome them.
This is the main point of this work: to propose a new calibration approach allowing smooth operation of non-modulated PWFS.

We based our approach on two previous works: \cite{2005ApOpt..44...60C} and \cite{2012SPIE.8447E..2BP}.
In the former, it is shown that the residual turbulence on the PWFS has an effect similar to the one of the modulation: it affects the sensitivity of the wavefront sensor (WFS).
This means, on \textcolor{black}{the one hand, that the the effective sensitivity is lowered when} the residual level increases (\textcolor{black}{this is in the right direction because the linearity of the PWFS increases when it is required to be larger}) and on \textcolor{black}{the other hand, a calibration taking into account this effect would help improve the performance of the system}.
In \cite{2012SPIE.8447E..2BP}, the authors present an on-sky calibration, in partial correction regime, \textcolor{black}{that is representative of the effective operation point of the sensor}.
\textcolor{black}{The goal of this work is to reproduce synthetically the effect of this residual turbulence to include it in the interaction matrix model}.
\textcolor{black}{The method to generate such matrices is called SIMPC for synthetic interaction matrix model in partial correction. In this paper, its characteristics are compared to a classical modulated PWFS for the calibration.}

The work is organized as follows.
\textcolor{black}{The parameters used in the simulations are summarized in Table \ref{Tab:params} and detailed in Appendix \ref{sec:simParams}.}
In Sect. \ref{sec:howto} we present the approach proposed to calibrate the PWFS (this is a general approach valid for both modulated and non-modulated configurations), then we show the sensitivity, linear range, and flat wavefront signal of the modulated and non-modulated PWFS (in Sects. \ref{sec:sens}, \ref{sec:linearity}, and \ref{sec:slopesNull} respectively \textcolor{black}{-- for the case of an 8m-class telescope}).
\textcolor{black}{Sections \ref{sec:results} and \ref{sec:elt} present full end-to-end simulations to assess the performance of the method, respectively, with 8m- and 39m-class telescopes.}
Finally, conclusions are presented in Sec. \ref{sec:conclusion}.
%
\begin{table*}
\caption{Simulation parameters. Further details can be found in Appendix \ref{sec:simParams}. Note: ELT is Extremely Large Telescope \textcolor{black}{\citep{2020SPIE11445E..1ET}}, DM is deformable mirror, and KL is Karhunen-Lo{\'e}ve \citep{Wang:78}.}
\label{Tab:params}
\begin{center}
\begin{small}
        \begin{tabular}{lcc}
                \hline
                \textbf{} & \textbf{8m-class telescope} & \textbf{39m-class telescope (ELT)}\\
                \hline
        \multicolumn{3}{l}{\textbf{Telescope}}\\
                \hline
        pupil diam. [m] & 8 & 39\\
        pupil samp. [pix] & 220 & 512\\
        obs. ratio [\%] & 11.1 & 28.3\\
        zenith angle [deg] & 30 & 30\\
                \hline
        \multicolumn{3}{l}{\textbf{Turbulence}}\\
                \hline
        \multirow{2}{*}{profile} & 4 layers  & 35 layers ``median'' profile\\
            & \citep[see][Table 10]{2014ExA....37..503A} & \citep[see][]{2013aoel.confE..89S}\\
        outer scale [m] & 40 & 25\\
                \hline
        \multicolumn{3}{l}{\textbf{WFS}}\\
                \hline
        no. sub-apertures & 40$\times$40 (max) & 90$\times$90 \\
        sub-pupils separation [pix] & 48 & 108\\
        field-of-View ($\phi$) [arcsec] & 2.1 & 2.1\\
        sensing wavelength [nm] & 750 & 798\\
        0-mag. det. flux [Gph/m$^2$/s] & 6.37 & 1.87\\
        photon noise & yes & yes \\
        excess noise & yes & yes \\
        read-out noise [e-] & 0.3 & 0.3\\
                \hline
        \multicolumn{3}{l}{\textbf{DM}}\\
                \hline
        conjugation altitude [m] & 0 & 0\\
        no. actuators & 672  & - \\
        no. modes & 630 &  4098\\
        modal basis & tip and tilt plus 628 KL modes & tip and tilt plus 4096 KL modes \\
                \hline
        \multicolumn{3}{l}{\textbf{Control}}\\
                \hline
        approach & modal & modal\\
        reconstruction matrix & least square solution & minimum mean square error estimator \\
        temporal control & leaky integrators  & leaky integrators\\
        framerate [Hz] & 1700 -- 250 & 1000 \\
        delay [ms] & 2 -- 5 & 3\\
                \hline
        \end{tabular}
\end{small}
\end{center}
\end{table*}

\section{How to use this approach}\label{sec:howto}

\textcolor{black}{This section is focused on the SIMPC approach to calibrate a non-modulated PWFS. As noted in the introduction, the method is based on two key-points: (i) that AO residuals affect the sensitivity as a modulation would \citep{2005ApOpt..44...60C} and (ii) that the system can be calibrated around this working point \citep{2012SPIE.8447E..2BP}.}

The effects of partial correction on PWFS, the so-called optical gain, has been studied in a number of papers \citep{2015aoel.confE..36E,2018SPIE10703E..20D,2019A&A...629A.107D,2020A&A...636A..88E,2020A&A...644A...6C,2021A&A...649A..70C}.
It is a well-know feature of the PWFS and it decreases for smaller modulation radius \citep[see][Fig. 5]{2018SPIE10703E..20D}.
Optical gain is a loss of sensitivity due to partial correction and its effect on the closed loop gain is a model error between the interaction matrix, calibrated \textcolor{black}{under} controlled or diffraction-limited conditions and the actual interaction matrix in partial correction regime.
\textcolor{black}{Thus}, while we cannot recover the sensitivity of the PWFS working in diffraction-limited conditions, we can \textcolor{black}{eliminate} this model error (or greatly reduce it) by calibrating the system in a condition as \textcolor{black}{close} as possible \textcolor{black}{to the one obtained} during operation.
\textcolor{black}{Although this is a general approach, this is particularly critical for non-modulated PWFS, as the model error is important between operation around a diffraction-limited PSF and on-sky operations impacted by AO residuals.}
We note that this approach is not equivalent to the modal optical gain compensation proposed in \cite{2018SPIE10703E..20D}, \cite{Chambouleyron2021}, and \cite{2021MNRAS.508.1745A} because all these methods are based on the approximation that the modal optical gain matrix is diagonal, while the \textcolor{black}{SIMPC also accounts} for the extra-diagonal elements of this matrix (see Sect. \ref{sec:sens}).
\textcolor{black}{The SIMPC method consists then in calibrating the interaction matrix around typical AO residuals, similarly to what is presented in \cite{2012SPIE.8447E..2BP}, with the difference that everything is done synthetically and does not requires on-sky access.}
Another difference, which is not explicitly discussed in \cite{2012SPIE.8447E..2BP}, is that the partial correction used for the calibration must \textcolor{black}{also include} any possible static or quasi-static non-common path aberration between science and PWFS paths that the PWFS would see to \textcolor{black}{ensure} a good correction of the science field.
\cite{2018MNRAS.481.2829H} has shown a synthetic reconstruction matrix \textcolor{black}{can be computed for the PWFS and used} on sky for an 8m telescope AO system and \textcolor{black}{this strategy is the baseline} on AOF and ERIS at VLT \citep[see][]{2018SPIE10703E..1GO,2022SPIE12185E..08R}.
\textcolor{black}{We note that this does not resolve all effects related to optical gains} because the SIMPC will be computed for a given partial correction (or a set of partial corrections) and cannot account for \textcolor{black}{all the} working conditions an AO system will face during on-sky operations.
\textcolor{black}{Nevertheless, the SIMPC provides an important advantage as the model errors between the calibration working point and on-sky working point will be much smaller than if the calibration is made under diffraction limit conditions.}
%
\begin{figure}
    \begin{center}
    \begin{tabular}{c}
        \includegraphics[width=0.4\textwidth]{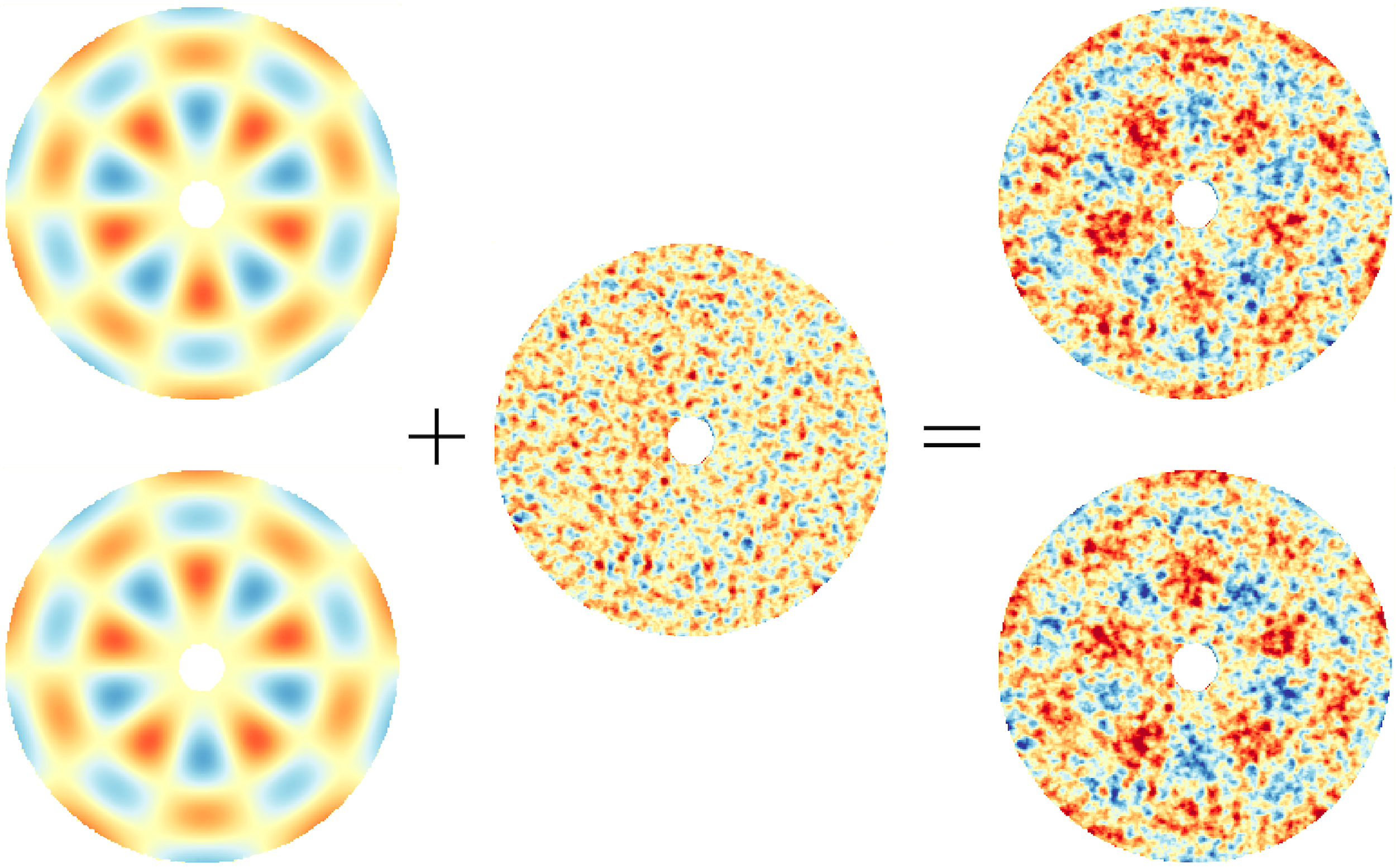}
    \end{tabular}
    \end{center}
         \caption{\label{fig:AberPushPull} Example of the phase used to run a single step of the push-pull calibration with partial correction \textcolor{black}{(for the 8m-class telescope case and 0.2 m sup-aperture size)}: on the left part mode no. 50 of the modal base with positive (top, push) and negative (bottom, pull) sign, on the central part a random realization of the partial correction \textcolor{black}{(i.e., turbulence residual),} and on the right part the combination of mode and partial correction. The wavefront root mean square (RMS) of the mode is 40nm and the wavefront RMS of the random realization of the partial correction is 46nm (corresponding to the residual correction level expected for a seeing of 0.4arcsec). We note that these RMS values are selected for display purpose, in general, the amplitude of the push-pull (i.e., the wavefront RMS of the mode) is much smaller than the RMS of the partial correction.
     }
\end{figure}

\begin{figure}
    \begin{center}
    \begin{tabular}{c}
        \includegraphics[width=0.45\textwidth]{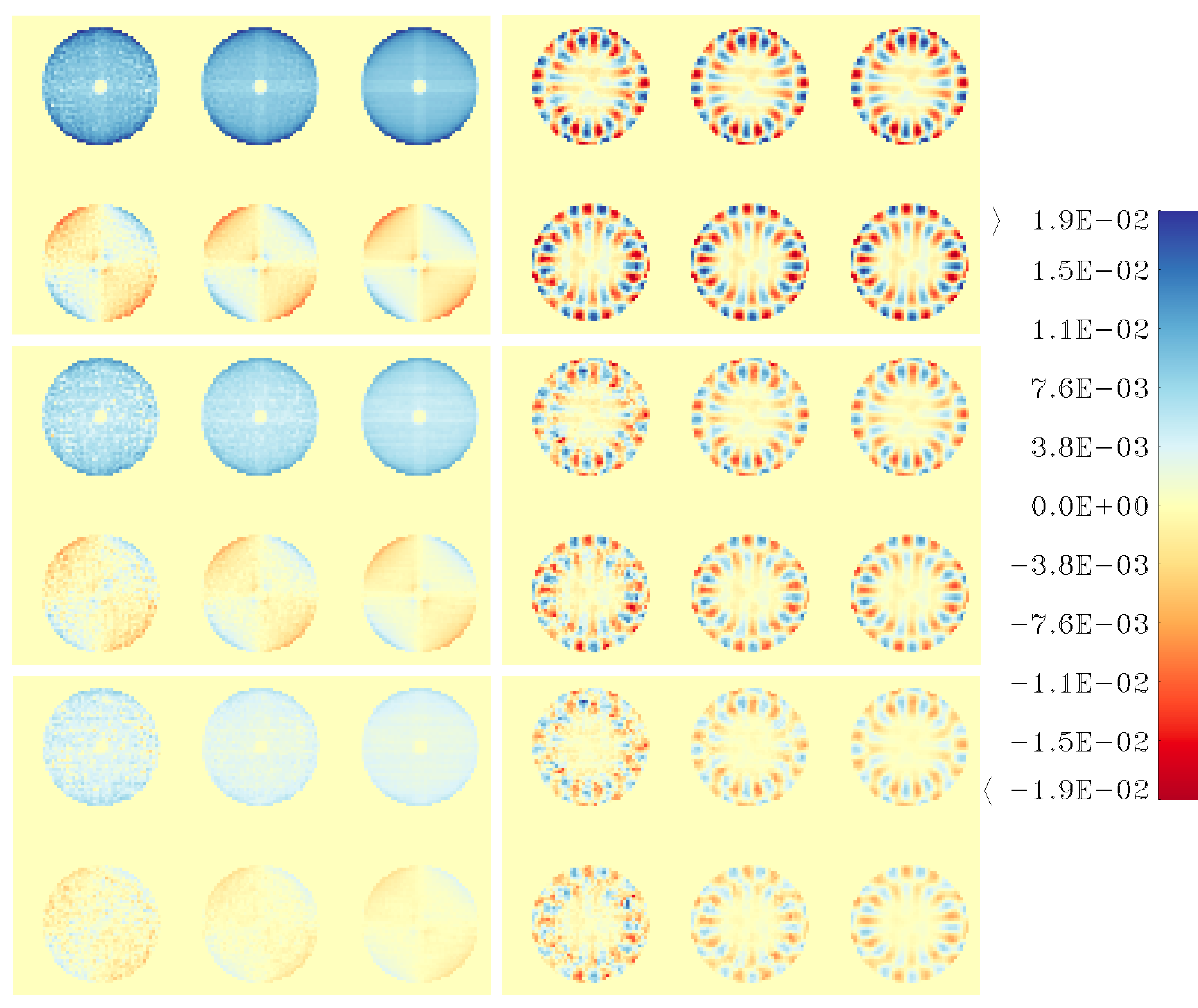}
    \end{tabular}
    \end{center}
         \caption{\label{fig:im_singals_cycles} \textcolor{black}{Tip (left) and mode no. 100 (right)} signals (x and y) from interaction matrices with different numbers of partial correction realizations \textcolor{black}{(from left to right): 1, 8, and 70}. Top: Partial correction for a seeing of 0.4arcsec, middle, partial correction for a seeing of 0.8arcsec, bottom, partial correction for a seeing of 1.2arcsec. Note: the signals amplitude is reduced when seeing increases. Refers to the sub-aperture size of 0.2 m. \textcolor{black}{8m-class telescope case.}}
\end{figure}

\begin{figure}
    \begin{center}
    \begin{tabular}{c}
        \includegraphics[width=0.4\textwidth]{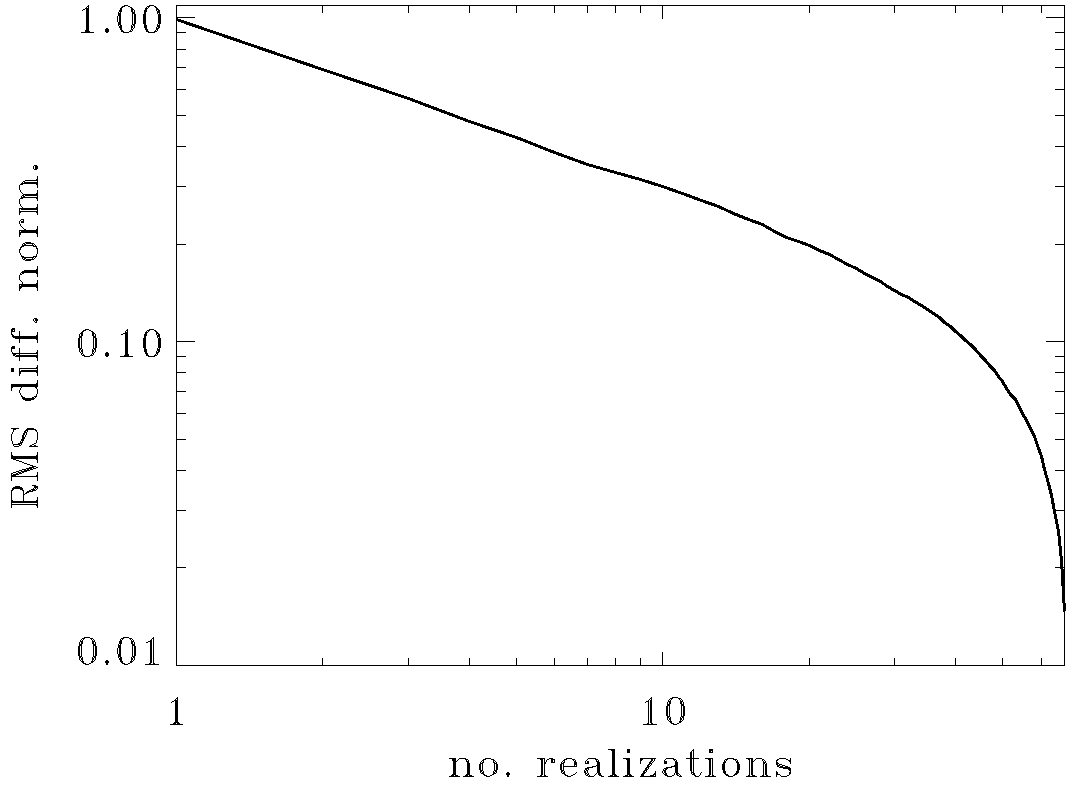}
    \end{tabular}
    \end{center}
         \caption{\label{fig:imdiff} RMS of the difference between the IM computed with $x+1$ and $x$ realizations of partial correction normalized by the RMS of the IM computed with $N$=70 realizations. Seeing is 1.4 arcsec and sub-aperture size is 0.2 m for the 8m-class telescope case.}
\end{figure}

In practice, the synthetic calibration is done mode-by-mode, with a push-pull signal repeated $N$ times, each time with a different \textcolor{black}{(ideally statistically independent)} realization of the desired partial correction level (an example for one mode and a casual realization of partial correction is shown in Fig. \ref{fig:AberPushPull}).
So, the i-th column of the interaction matrix, $D,$ will be:

\begin{equation}\label{eq:im}
    D_{i} = \frac{1}{2N} \sum_{j=1}^{\textcolor{black}{2N}} \frac{s_{j}}{(-1)^{2(j+1)}a_{i}} \; ,
\end{equation}
where $s$ is the slope vector \textcolor{black}{(whose description is given in Appendix \ref{sec:simParams})} and $a$ is the absolute value of the amplitude of the mode applied.
In principle, this can be done also considering $M$ different levels of partial correction for a total of $NM$ \textcolor{black}{repetitions}.
In this case, Eq. (\ref{eq:im}) becomes:

\begin{equation}\label{eq:im2}
    D_{i} = \frac{1}{2NM} \sum_{k=1}^{\textcolor{black}{M}} \sum_{j=1}^{\textcolor{black}{2N}} \frac{s_{(j,k)}}{(-1)^{2(j+1)}\textcolor{black}{a_{i}}} \; .
\end{equation}
\textcolor{black}{An example of} the effect of the number of \textcolor{black}{repetitions} is shown in Fig. \ref{fig:im_singals_cycles}: a single realization gives noisy signals and only with several tens of realizations the convergence is reached for the worst seeing considered.
In particular, for a case with sub-aperture sizes of 0.2, 0.4, 0.6, and 0.8 m, we will use $N$=70, 100, 200, and 200, respectively, in the rest of the paper.
The \textcolor{black}{RMS} of the difference between the IM computed with $N$ and $N-1$ realizations is <0.02 of the RMS of IM computed with $N$ with the values of $N$ mentioned above for a seeing of 1.4 arcsec. \textcolor{black}{An example of this convergence is shown in Fig. \ref{fig:imdiff} for the sub-aperture size of 0.2 m}.
\textcolor{black}{Note: the realizations we used were generated from a single simulation, we selected turbulence residual samples separated by 100 simulation steps (0.059s for the maximum framerate of 1700 Hz).
Adjacent samples are not statistically independent, they have correlation levels of about 10\%.}

\section{Sensitivity and optical gains}\label{sec:sens}
\begin{figure*}
    \begin{center}
    \begin{tabular}{c}
        \includegraphics[width=0.8\textwidth]{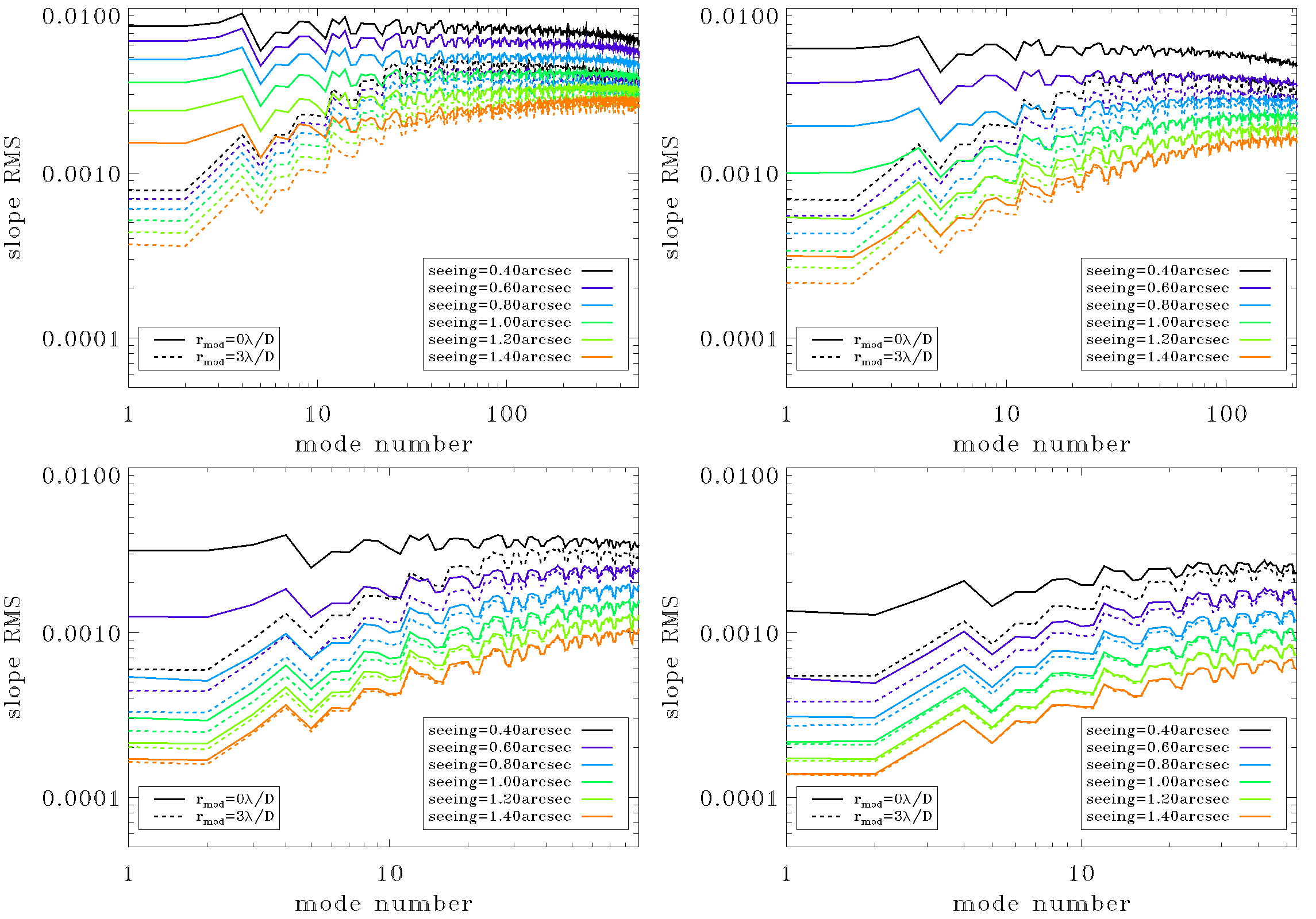}
    \end{tabular}
    \end{center}
         \caption{\label{fig:sens_b1_2_3_4} PWFS sensitivity in partial correction as a function of mode number and seeing value expressed as the RMS of the slope for an aberration with \textcolor{black}{a wavefront} RMS of 1nm. Mode 1 is tip, 2 is tilt and the other modes are Karhunen-Lo\`eve modes.
     Solid lines correspond to a non-modulated PWFS and dashed lines correspond to a PWFS with a modulation radius of 3$\lambda/D$. Top-left:\ Sub-aperture size of 0.2 m, top-right, sub-aperture size of 0.4 m. Bottom-left: Sub-aperture size of 0.6 m. Bottom-right:\ Sub-aperture size of 0.8 m for the \textcolor{black}{8m-class telescope case.}}
\end{figure*}
The main reason to use a non-modulated PWFS \textcolor{black}{is for the gain in sensitivity}.
\textcolor{black}{This is particularly true for low-order spatial frequencies as presented} in \cite{2021A&A...650L...8C}.
This \textcolor{black}{higher} sensitivity is relevant when dealing with differential piston sensing as shown, for example, in \cite{2022JATIS...8b1502E} and \cite{2022SPIE12185E..59L}.

The sensitivity in partial correction as a function of mode number and seeing value expressed as the \textcolor{black}{RMS}, $x_{\mathrm{RMS}}$, of the \textcolor{black}{WFS signal (also known as slope)} for \textcolor{black}{an aberration} with \textcolor{black}{a wavefront} RMS of 1nm \textcolor{black}{is reported in Fig. \ref{fig:sens_b1_2_3_4}}.
The i-th element of the $x_{\mathrm{RMS}}$ vector is:

\begin{equation}
    x_{\mathrm{RMS}}(i) = \sqrt{\frac{1}{N_{sl}}\sum_{j=1}^{N_{sl}}D_{i}^2(j)} \; ,
\end{equation}
where $D$ is the interaction matrix and $N_{sl}$ is the number of elements of the wavefront sensor (WFS) signal vector.
Here, we are considering a partial correction given by a PWFS working in high-flux conditions. \textcolor{black}{As can be deduced from Fig. \ref{fig:sens_b1_2_3_4}, there are 48 sets of turbulence residuals (i.e., with a partial correction) used to calculate the SIMPCs: their RMS and the RMS of the WFS signals calculated along with them are given in Tables \ref{Tab:turbResRMS} and \ref{Tab:slopeRMS}.}

\textcolor{black}{The result shown in Fig. \ref{fig:sens_b1_2_3_4} confirms the qualitative analysis provided} in the first part of this section, the non-modulated PWFS is generally more sensitive than a modulated PWFS \textcolor{black}{\citep[we chose a modulation radius of 3$\lambda/D$ for the comparison because is the current value used for SOUL at LBT, see][]{Pinna2019}}: the sensitivity gain is larger for smaller sub-aperture size, where the residual is lower because the fitting error is lower (a larger number of modes is corrected) and is larger for lower order modes, in particular, for the first two, tip and tilt.
The gain is between a factor 4 (bad seeing) and 10 (good seeing) for tip and tilt and a factor 1.5 on mode 500 when the sub-aperture size is 0.2 m.
The gain decreases when the sub-aperture size is reduced: it becomes 1 for bad seeing and a sub-aperture of 0.8 m. 

\begin{figure}
    \begin{center}
    \begin{tabular}{c}
        \includegraphics[width=0.425\textwidth]{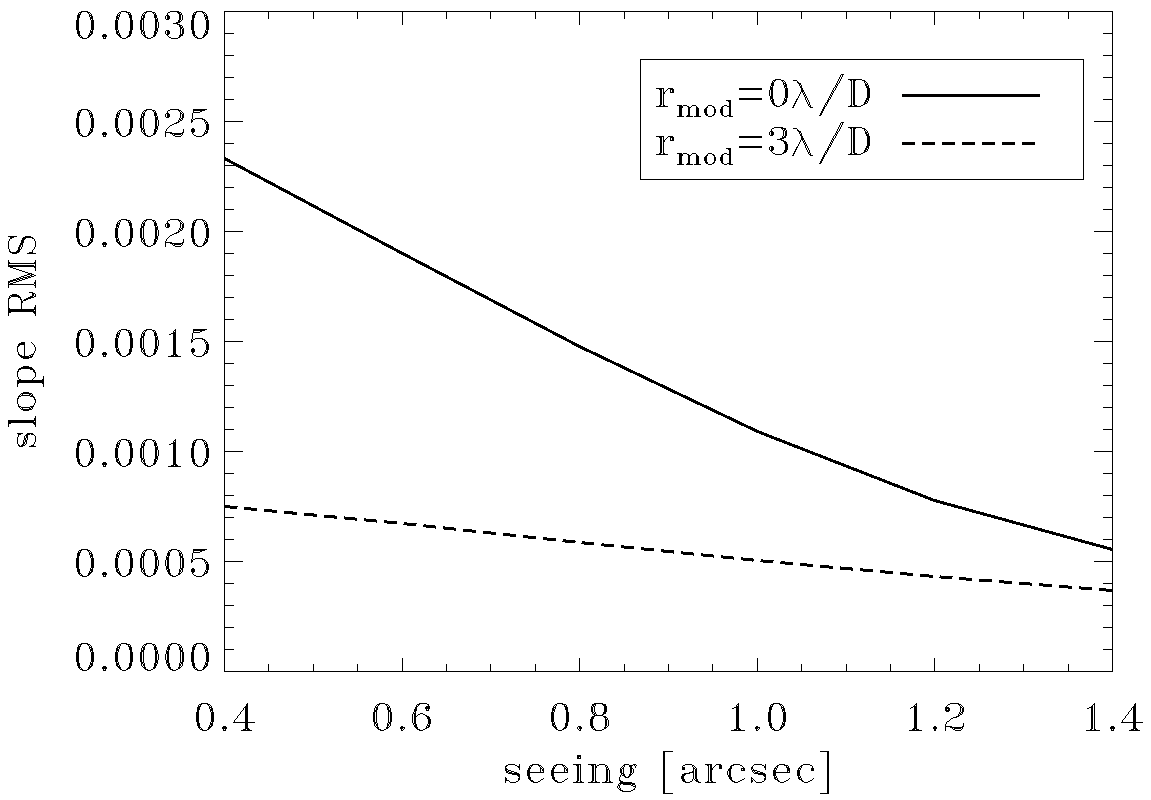}
    \end{tabular}
    \end{center}
         \caption{\label{fig:sens_pist} PWFS sensitivity to a differential piston mode in partial correction as a function of seeing value expressed as the RMS of the slope for an aberration with \textcolor{black}{a wavefront} RMS of 1nm. The differential piston mode is the one produced by one petal of a six-petal configuration with no gap. Solid line correspond to a non-modulated PWFS and dashed line correspond to a PWFS with a modulation radius of 3$\lambda/D$. Sub-aperture size is 0.2 m. \textcolor{black}{Refers to the 8m-class telescope case.}}
\end{figure}

\begin{figure}
    \begin{center}
    \begin{tabular}{c}
        \includegraphics[width=0.2\textwidth]{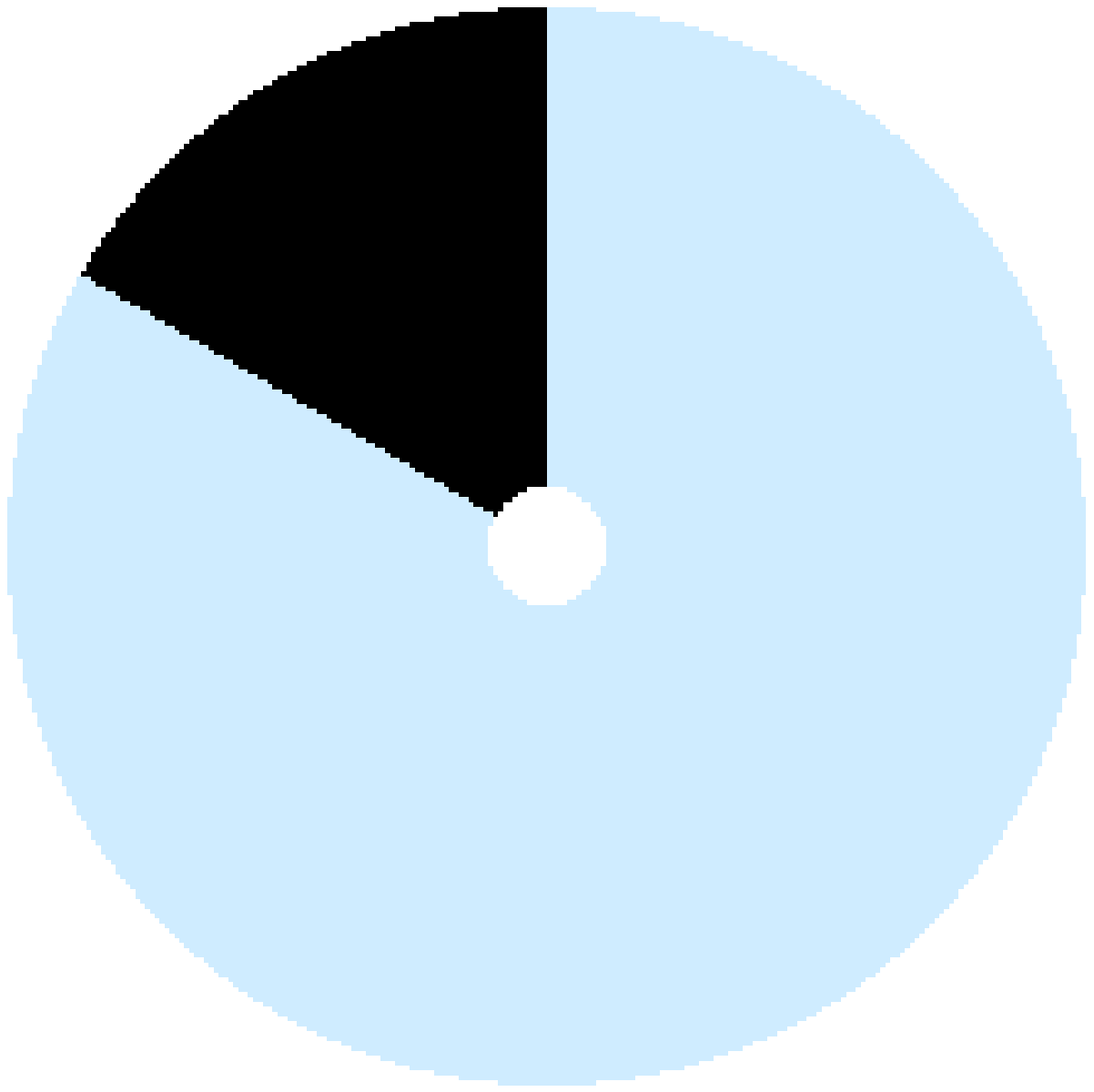}
    \end{tabular}
    \end{center}
         \caption{\label{fig:petal_mode} Differential piston mode \textcolor{black}{in black} (also \textcolor{black}{known} as petal mode) produced by one segment (petal) of a six-segment\ (petals) configuration with no gap. The pupil is shown in light blue. Refers to the 8m-class telescope case.}
\end{figure}

\textcolor{black}{Since this work is particularly relevant for pupil fragmentation sensing, Fig. \ref{fig:sens_pist} gives} the sensitivity to a differential piston mode in partial correction as a function seeing value expressed as the RMS of the \textcolor{black}{WFS signal} for an aberration with \textcolor{black}{a wavefront} RMS of 1nm (an infinite \textcolor{black}{signal-to-noise} ratio, S/N).
We note that the piston mode is the one produced by one petal of a six-petal configuration with no gap (see Fig. \ref{fig:petal_mode}).
\textcolor{black}{This ability to better see phase jumps is of particular interest for the next generation of AO instruments,  as well as for existing systems that may be impacted by thermal effects of the spiders \citep{2015aoel.confE...9S,2018A&A...615A..34W,2022A&A...665A.158P}.}
%
\begin{table}
\caption{RMS of turbulence residuals (i.e. partial correction) in nm for a bright star condition.
PWFS with and without modulation and different seeing and sub-aperture size values are considered.
These values are computed on the sets of turbulence residual samples used to compute the SIMPCs (see Sect. \ref{sec:howto}). Refers to the 8m-telescope case.}
\label{Tab:turbResRMS}
\begin{small}
\begin{center}
        \begin{tabular}{ccccccc}
                \hline
                \textbf{sub-ap.} & \multicolumn{6}{c}{\textbf{seeing [arcsec]}} \\
          \textbf{size [m]} & 0.4 & 0.6 & 0.8 & 1.0 & 1.2 & 1.4 \\
                \hline
        \multicolumn{7}{c}{\textbf{$r_{mod} = 0 \lambda/D$}}\\
            \hline
        0.2 & 48 & 67 & 88 & 111 & 137 & 164 \\
        0.4 & 73 & 107 & 143 & 164 & 219 & 261 \\
        0.6 & 116 & 174 & 240 & 303 & 365 & 428 \\
        0.8 & 178 & 270 & 360 & 456 & 551 & 660 \\
                \hline
        \multicolumn{7}{c}{\textbf{$r_{mod} = 3 \lambda/D$}}\\
            \hline
        0.2 & 48 & 67 & 84 & 103 & 122 & 144 \\
        0.4 & 76 & 107 & 138 & 175 & 214 & 254 \\
        0.6 & 122 & 177 & 234 & 292 & 350 & 416 \\
        0.8 & 188 & 272 & 360 & 450 & 551 & 659 \\
                \hline
        \end{tabular}
\end{center}
\end{small}
\end{table}
\begin{table}
\caption{RMS of the slopes \textcolor{black}{(WFS signals)} computed on the sets of turbulence residual (i.e., partial correction) samples used to compute the SIMPCs (see Sect. \ref{sec:howto}).
PWFS with and without modulation, infinite S/N, and different seeing and sub-aperture size values are considered.
Note: the turbulence residual has two opposite effects: it increases amplitude of signal to be measured (larger RMS) and it reduces the optical gain of the PWFS (smaller RMS, see Sect. \ref{sec:sens}). Refers to the 8m-telescope case.}
\label{Tab:slopeRMS}
\begin{small}
\begin{center}
        \begin{tabular}{ccccccc}
                \hline
                \textbf{sub-ap.} & \multicolumn{6}{c}{\textbf{seeing [arcsec]}} \\
          \textbf{size [m]} & 0.4 & 0.6 & 0.8 & 1.0 & 1.2 & 1.4 \\
                \hline
        \multicolumn{7}{c}{\textbf{$r_{mod} = 0 \lambda/D$}}\\
            \hline
        0.2 & 0.220 & 0.267 & 0.296 & 0.315 & 0.329 & 0.337 \\
        0.4 & 0.243 & 0.270 & 0.284 & 0.291 & 0.291 & 0.287 \\
        0.6 & 0.303 & 0.314 & 0.311 & 0.302 & 0.292 & 0.283 \\
        0.8 & 0.343 & 0.332 & 0.311 & 0.299 & 0.290 & 0.283 \\
                \hline
        \multicolumn{7}{c}{\textbf{$r_{mod} = 3 \lambda/D$}}\\
            \hline
        0.2 & 0.221 & 0.231 & 0.250 & 0.271 & 0.292 & 0.306 \\
        0.4 & 0.215 & 0.233 & 0.253 & 0.265 & 0.270 & 0.270 \\
        0.6 & 0.255 & 0.277 & 0.283 & 0.283 & 0.279 & 0.274 \\
        0.8 & 0.285 & 0.296 & 0.292 & 0.287 & 0.282 & 0.278 \\
                \hline
        \end{tabular}
\end{center}
\end{small}
\end{table}
%
%
\begin{figure}
    \begin{center}
    \begin{tabular}{c}
        \includegraphics[width=0.425\textwidth]{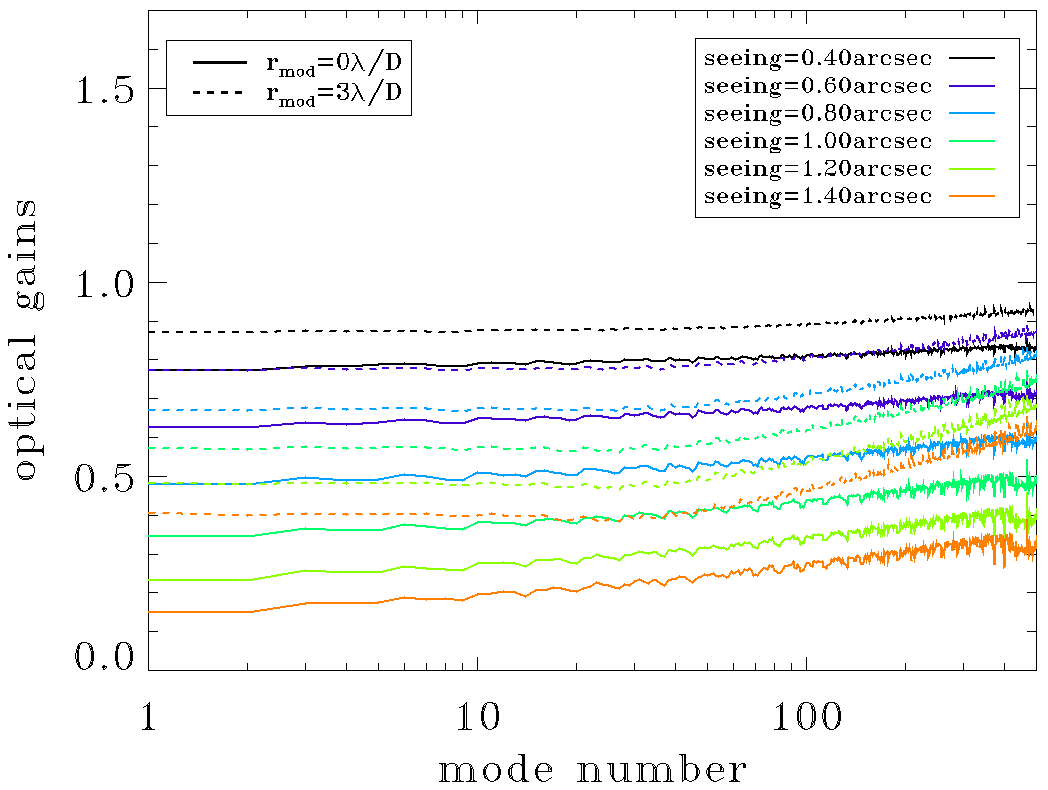}
    \end{tabular}
    \end{center}
         \caption{\label{fig:optical_gain} Modal optical gain for the PWFS calibrated in diffraction limited conditions. Mode 1 is tip, 2 is tilt, and the other modes are Karhunen-Lo\`eve modes.
     Solid lines correspond to a non-modulated PWFS and dashed lines correspond to a PWFS with a modulation radius of 3$\lambda/D$. Refers to the 8m-class telescope case.}
\end{figure}
\begin{figure}
    \begin{center}
    \begin{tabular}{c}
        \includegraphics[width=0.4\textwidth]{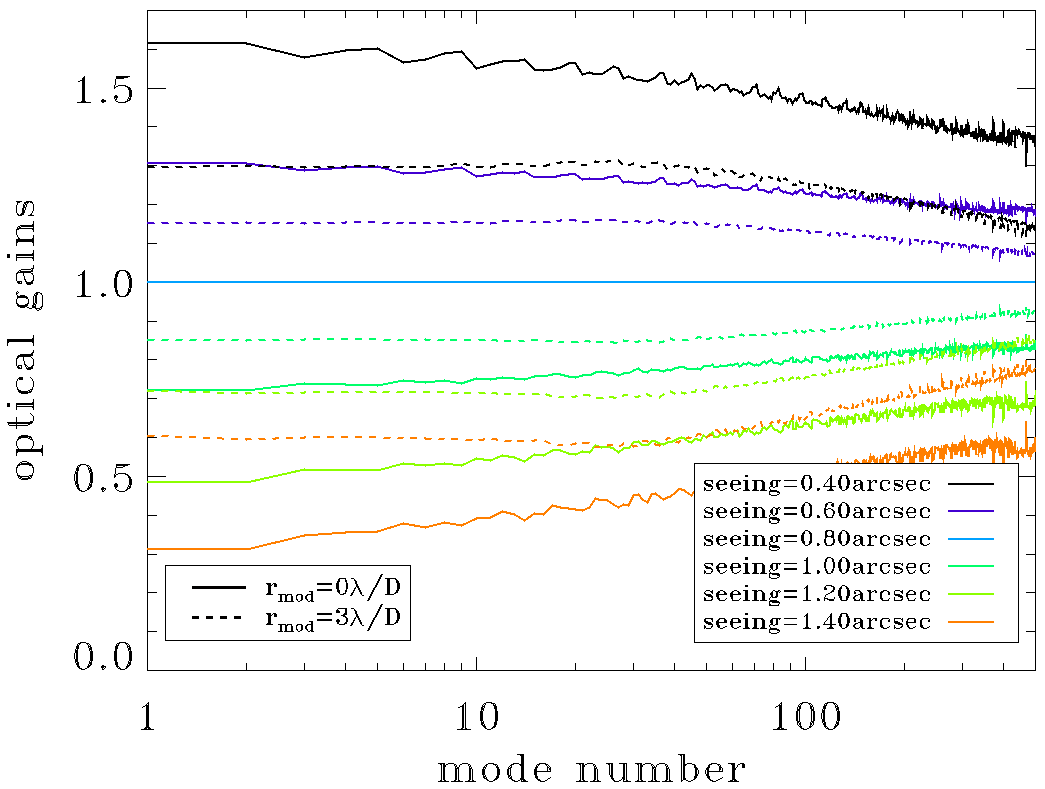}
    \end{tabular}
    \end{center}
         \caption{\label{fig:optical_gain_partial} Modal optical gains for the PWFS calibrated in partial correction with a seeing value of 0.8arcsec. Mode 1 is tip, 2 is tilt, and the other modes are Karhunen-Lo\`eve modes.
     Solid lines correspond to a non-modulated PWFS and dashed lines correspond to a PWFS with a modulation radius of 3$\lambda/D$. \textcolor{black}{Refers to the 8m-class telescope case.}}
\end{figure}
\begin{figure}
    \begin{center}
    \begin{tabular}{c}
        \includegraphics[width=0.4\textwidth]{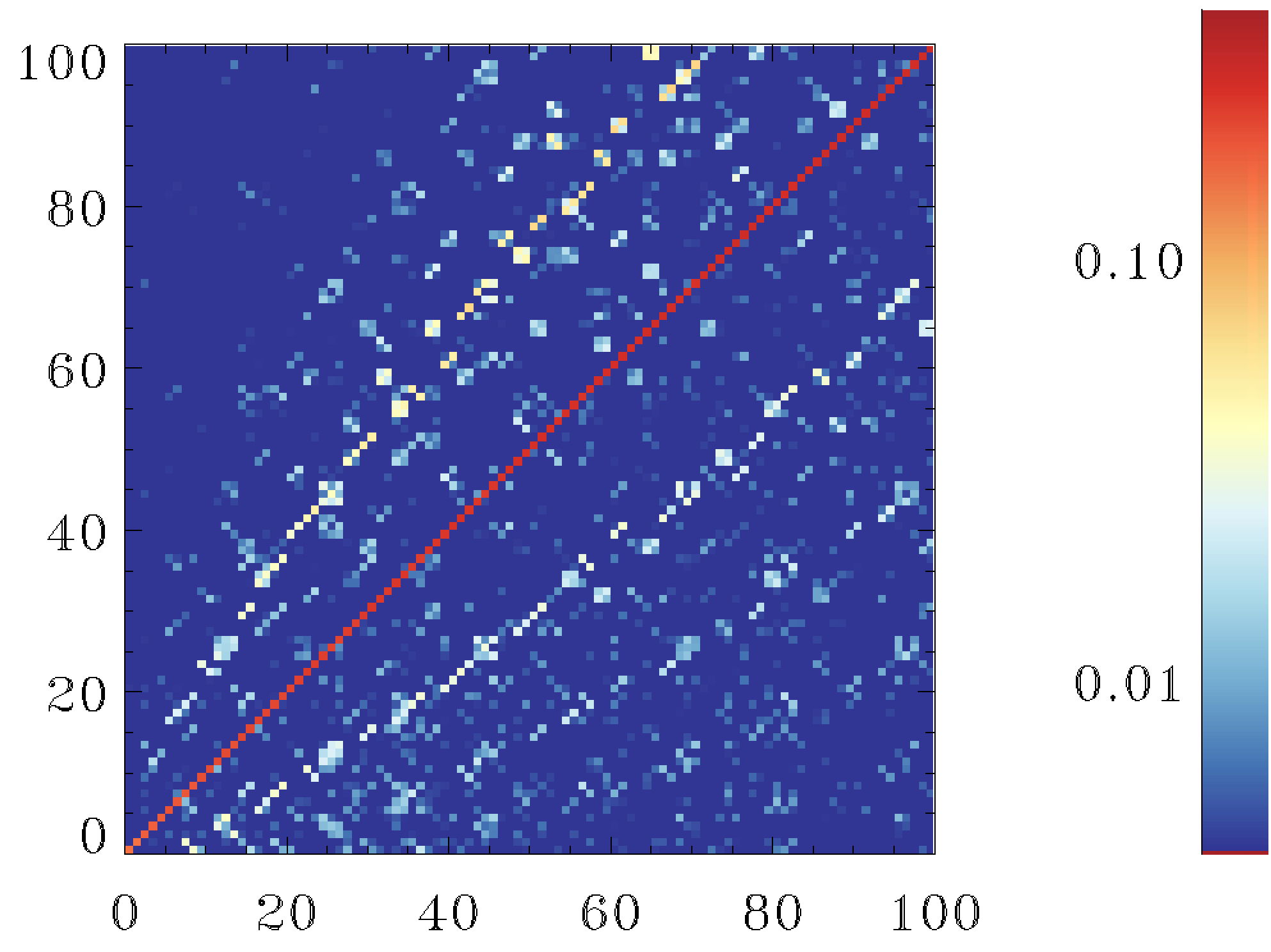}
    \end{tabular}
    \end{center}
         \caption{\label{fig:RMxIMex} Results of the product between reconstruction matrix in diffraction limited condition and SIMPC for seeing of 1.4arcsec, without modulation and with sub-aperture size of 0.2 m. This is the so called modal optical gain matrix. An absolute value is used and only the first 100 $\times$ 100 elements of the matrix is shown for clarity reasons. It can be easily seen that that the matrix is not diagonal. The diagonal of this matrix is shown in Fig. \ref{fig:optical_gain} (orange solid line). \textcolor{black}{Refers to the 8m-class telescope case.}}
\end{figure}

\textcolor{black}{We defined the optical gains as the scalar coefficients expressing the change in PWFS sensitivity due to partial correction}.
\textcolor{black}{An example is shown} in Fig. \ref{fig:optical_gain} for a PWFS calibrated in diffraction-limited conditions.

These gains must be compensated \textcolor{black}{to provide the best performance but as well} to get a proper non-common path correction, as shown in \cite{2015aoel.confE..36E}, \cite{2018SPIE10703E..20D}, \cite{2020A&A...636A..88E}, \cite{2020A&A...644A...6C}, and \cite{Chambouleyron2021}.
When the PWFS calibration takes into account a partial correction equivalent to \textcolor{black}{that found in operation,} the optical gains are \textcolor{black}{equal to 1; however, if there is} a discrepancy between the residual level in calibration and in operation, the optical gains are different from 1 (\textcolor{black}{may also be} greater than 1).
A summary of this effect is shown in Fig. \ref{fig:optical_gain_partial}.
\textcolor{black}{Interestingly,} the gains are closer to 1 than the diffraction-limited case and the \textcolor{black}{mode-by-mode difference is small}.
Here, we \textcolor{black}{consider} the diagonal approximation, but we note that (as \textcolor{black}{mentioned earlier in this paper}) the modal optical gains matrix is a full matrix (as can be seen in Fig. \ref{fig:RMxIMex}, where, an example of the product between reconstruction matrix in diffraction-limited condition and SIMPC is shown) and not a diagonal matrix.

\section{Linearity}\label{sec:linearity}

While the sensitivity of an non-modulated PWFS is \textcolor{black}{greater than that of} a modulated PWFS, the linearity of a non-modulated PWFS is reduced\footnote{Except for differential piston, because the response for both non-modulated and modulated PWFS to such aberrations is sinusoidal as it is shown in \cite{2002ESOC...58..161E}.} and this is still valid even \textcolor{black}{when using the SIMPC approach}.
\textcolor{black}{The model error can be reduced but the linearity can not be changed when operating in a non-null working point\footnote{An exception is exotic approaches such as the ones presented in \cite{2018JOSAA..35..594F}, \cite{2018ApOpt..57.8790H} and \cite{2023arXiv230509805A}.}.}
The linearity is improved by the partial correction as reported in \cite{2005ApOpt..44...60C}, but \textcolor{black}{there is no} direct control on this.
Typically, in closed loop, the aberration to be sensed are proportional to the residual aberration.
Another point to consider is the fitting error: when the number of modes and actuators decreases, as it happens for SOUL with dimmer stars when the detector is binned \citep[see][]{10.1117/12.2234444,Pinna2019}, the fitting error increases, and higher level of residual reduces the PWFS sensitivity.
This means that it is less important to modulate the PWFS when few modes or actuators are corrected.
\textcolor{black}{We must then consider that higher sensitivity gives greater benefits} when the signal-to-noise ratio is \textcolor{black}{lower}, so, in general, a non-modulated PWFS is preferable on dimmer stars, where we would also expect \textcolor{black}{worse} correction and higher residual.

\begin{figure*}
    \begin{center}
    \begin{tabular}{c}
        \includegraphics[width=0.8\textwidth]{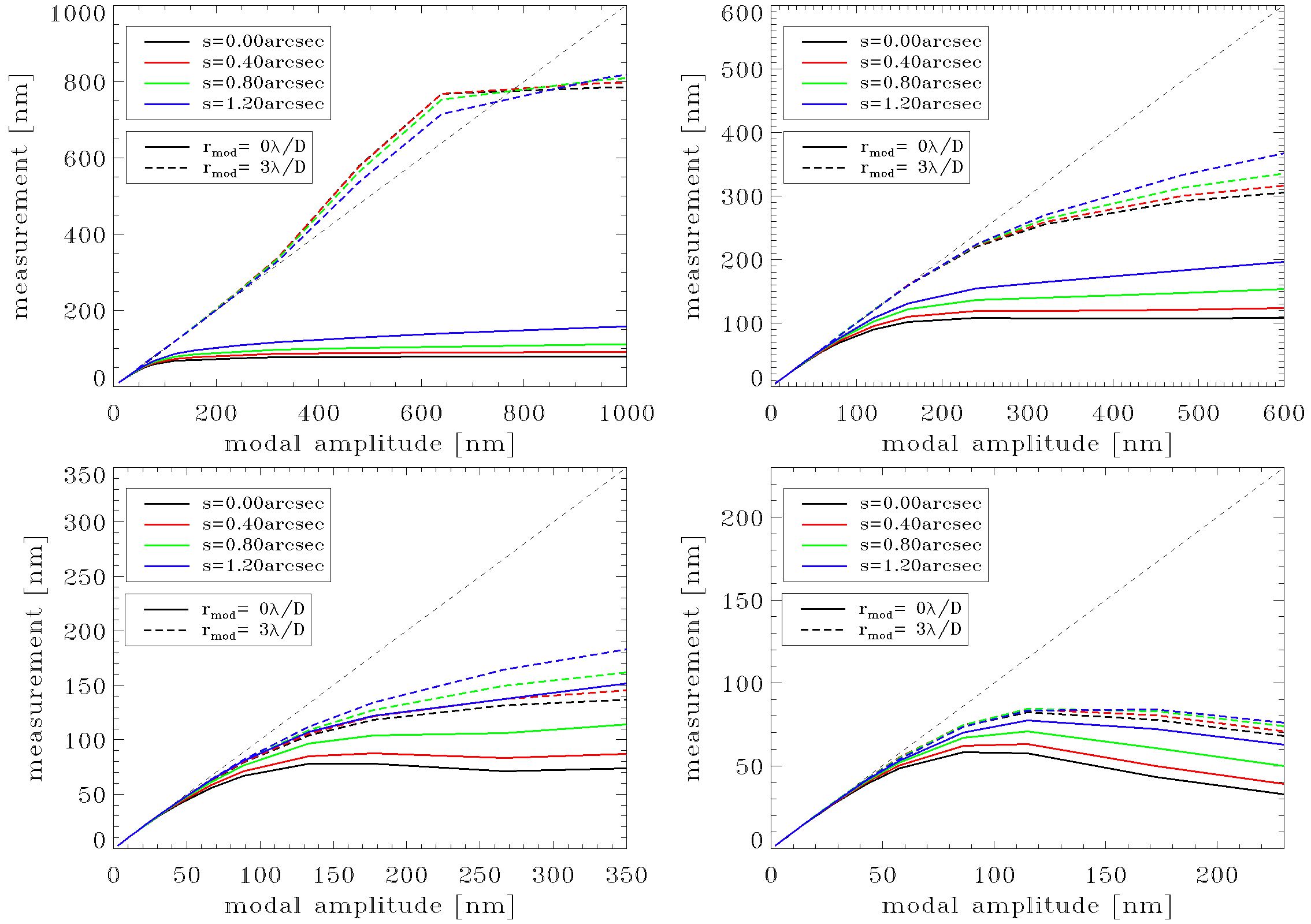}
    \end{tabular}
    \end{center}
         \caption{\label{fig:linear_b1_v2} Response of the PWFS to modes with different value of wavefront RMS and for different partial correction levels (s=0.00arcsec correspond to diffraction limited conditions).
     Top-left:\ Mode 0 (tip). Top-right:\ Mode 10. Bottom-left: Mode 100. Bottom-right:\ Mode 499.
     \textcolor{black}{Mode 0 is tip, 1 is tilt,} and the other modes are Karhunen-Lo\`eve modes.
     Solid lines correspond to a non-modulated PWFS and dashed lines correspond to a PWFS with a modulation radius of 3$\lambda/D$.
     Sub-aperture size is 0.2 m.
     Optical gain is corrected considering an aberration RMS of 20nm. Refers to the 8m-class telescope case.}
\end{figure*}

\begin{figure}
    \begin{center}
    \begin{tabular}{c}
        \includegraphics[width=0.4\textwidth]{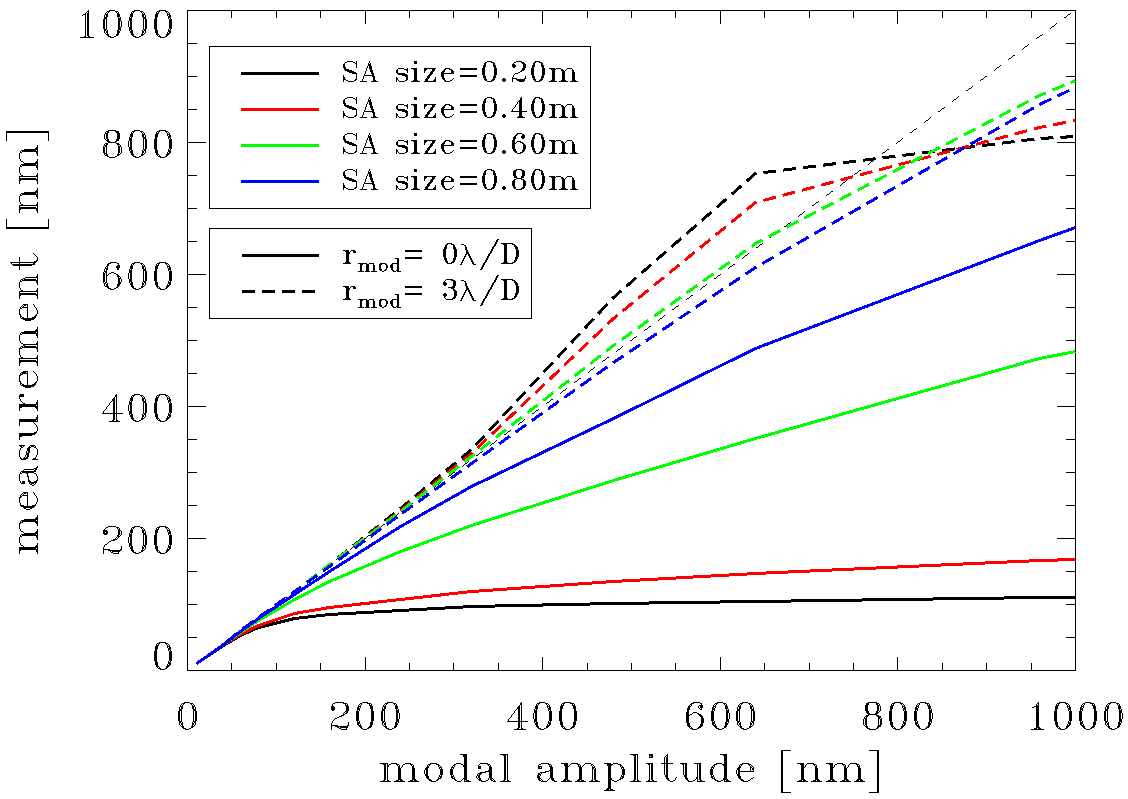}
    \end{tabular}
    \end{center}
         \caption{\label{fig:linear_b1_2_3_4} Response of the PWFS to \textcolor{black}{a tip} with different value of  wavefront RMS and for different sub-aperture size and for a seeing value of 0.8arcsec.
     Solid lines correspond to a non-modulated PWFS and dashed lines correspond to a PWFS with a modulation radius of 3$\lambda/D$. \textcolor{black}{Refers to the 8m-class telescope case.}}
\end{figure}

As for the previous section, a partial correction given by a PWFS working in high flux condition \textcolor{black}{is considered} (equivalent to the one presented in in Sect. \ref{sec:results} for an R magnitude of 8).
We report in Fig. \ref{fig:linear_b1_v2} the response of the PWFS to a set of modes with increasing wavefront RMS.
The correction condition range from diffraction limit to a seeing of 1.2 arcsec and the sub-aperture size is 0.2 m.
The modulated PWFS has a clear advantage on the first three modes considered, 0, 5, and 10.
\textcolor{black}{Specifically,  the modulated PWFS under} diffraction limited conditions is linear up to about 600nm on mode 0 (tip), 400nm on mode 5, and 200nm on mode 10.
The linearity increases with partial correction well beyond these value (see Fig. \ref{fig:linear_b1_v2}).
The non-modulated PWFS has a significant smaller linearity range, but it increases with partial correction, and the PWFS response on the lower order modes (a few tens) is monotonous \textcolor{black}{except for} the diffraction limited condition.
\textcolor{black}{Interestingly, by} increasing the sub-aperture size (\textcolor{black}{thus} the fitting error) the linear range greatly improves, as can be seen in Fig. \ref{fig:linear_b1_2_3_4}.
\section{Flat wavefront \textcolor{black}{WFS response}}\label{sec:slopesNull}

\begin{figure}
    \begin{center}
    \begin{tabular}{c}
        \includegraphics[width=0.475\textwidth]{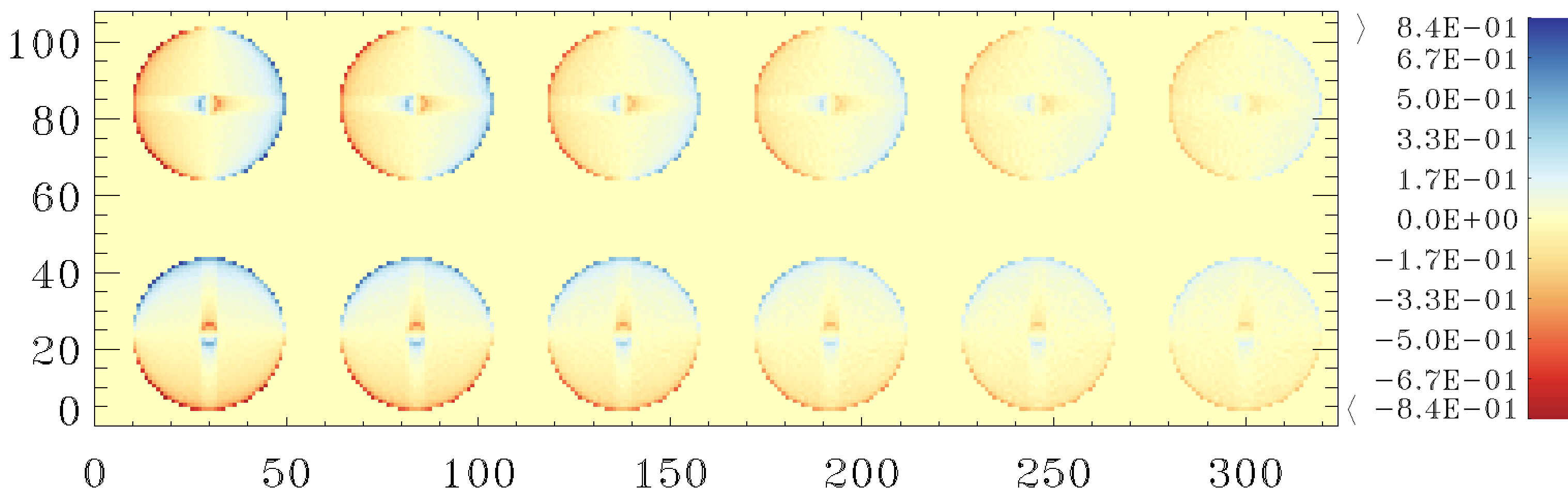}
    \end{tabular}
    \end{center}
         \caption{\label{fig:slopeNulls} Signals corresponding to a flat wavefront for a non-modulated PWFS with a 0.2 m sub-aperture and different level of partial correction corresponding to seeing values of 0.4, 0.6, 0.8, 1.0, 1.2, and 1.4 arcsec, respectively. Refers
to the 8m-class telescope case.}
\end{figure}

\textcolor{black}{This section focuses on the slope vector associated with a flat wavefront.}
For the PWFS, this \textcolor{black}{signal} is not a vector of zeros.
It is interesting to note that in case of a non-modulated PWFS the amplitude of the \textcolor{black}{signal} corresponding to a flat wavefront \textcolor{black}{is greater than the one} of a modulated PWFS, so it is more important to take \textcolor{black}{it} into account in this configuration.
For a sub-aperture size of 0.20 m, it is \textcolor{black}{a signal} corresponding to an aberration of about 20 nm, about a \textcolor{black}{factor 2 above} the value found for a \textcolor{black}{modulated} PWFS with modulation radius of 3$\lambda/D$.
Following a similar approach \textcolor{black}{as the SIMPC \citep[and as the one presented in][]{2022JATIS...8b1502E}}, we computed the flat wavefront slope vector as the average of a set of 500 slopes measured with a random residual that is given by the fitting error only: the wavefront spectrum has a zero amplitude up to the maximum spatial frequency that can be corrected by the AO system.
Examples of these signals are shown in Fig. \ref{fig:slopeNulls}.
We note that the slope intensity decreases with increasing partial correction (i.e., the seeing value); this means that when the seeing is changing, the flat wavefront slope reference should be changed as well \citep{2022JATIS...8b1502E}.

\section{Simulation results}\label{sec:results}

\subsection{\textcolor{black}{Simulation with no calibration error of the optical gain}}

\textcolor{black}{In this section, we present simulations where the value of seeing found during operation is the same value used for the calibration.
In this case, the optical gain should be automatically compensated by the SIMPC.}

\begin{figure}
    \begin{center}
    \begin{tabular}{c}
        \includegraphics[width=0.4\textwidth]{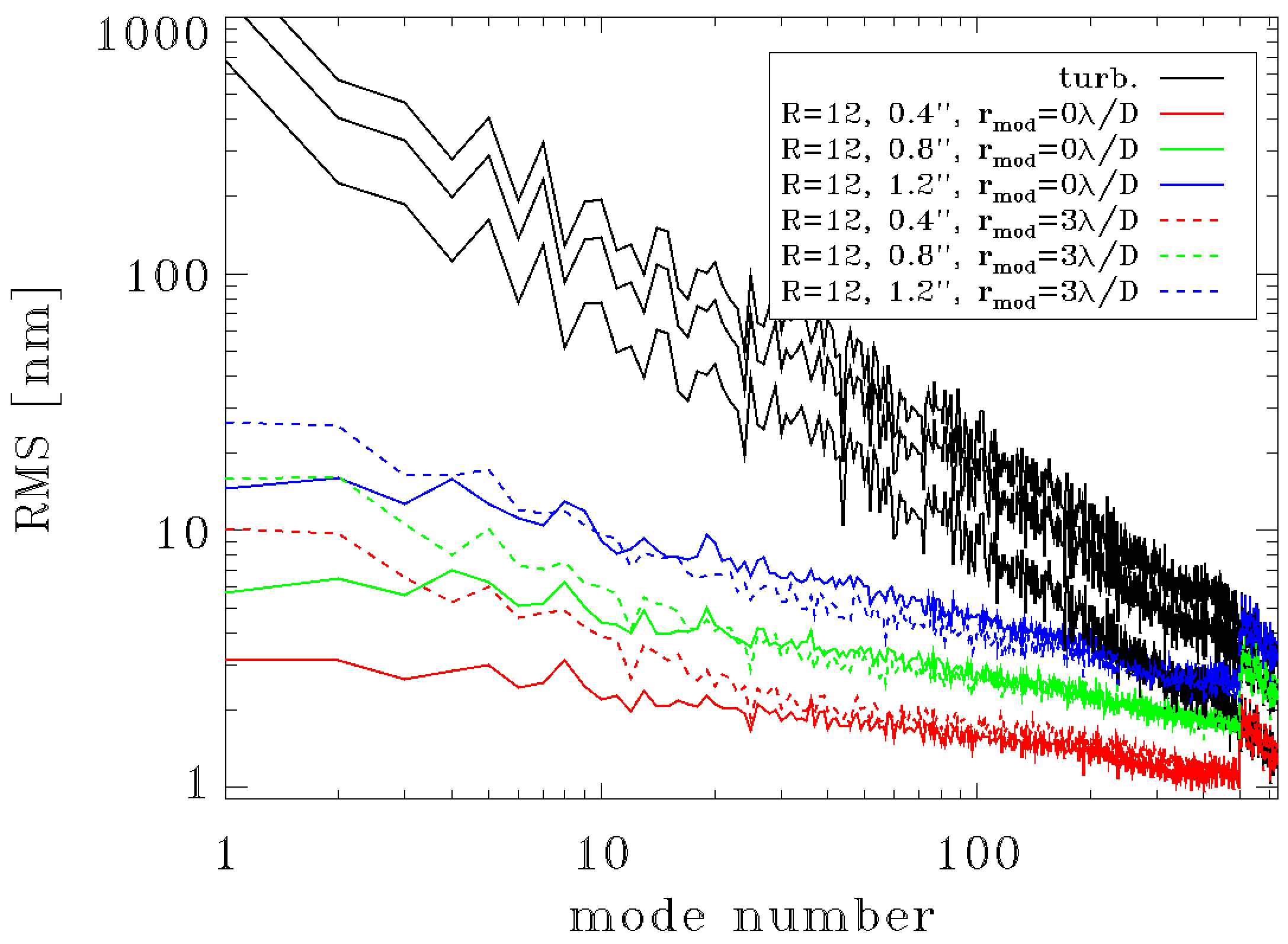}
    \end{tabular}
    \end{center}
         \caption{\label{fig:modal_plot_R12_bin1} Turbulence (black lines) and residuals (colored lines) modal RMS amplitudes for a NGS of magnitude R=12, different seeing values and for a non-modulated and modulated (modulation radius of 3$\lambda/D$, dashed lines) PWFS.
         Sub-aperture size is 0.2 m and 500 modes are corrected. \textcolor{black}{Refers to the 8m-class telescope case.}}
\end{figure}

We ran a set of simulations with PASSATA \citep[see][]{doi:10.1117/12.2233963} to evaluate the closed loop performance of the non-modulated PWFS \textcolor{black}{for the 8m-class telescope case}.
We used a control based on leaky integrators \citep[see][]{Agapito2019} and we optimized  the modal gain online, as described in \cite{2021MNRAS.508.1745A}.
\textcolor{black}{Additional information on the parameters used and on the details of the simulation can be found in Table \ref{Tab:params} and in Appendix \ref{sec:simParams}.}
We explored several magnitudes from R=8 to 16.5 and we used a system configuration analogous to the one of SOUL \citep[see][]{Pinna2019} where we binned the detector to deal with lower fluxes.
The summary of the results is presented in Table \ref{Tab:results}.
\textcolor{black}{Non-modulated PWFS shows better performance when observing faint objects under very good seeings.
Under these conditions, the limited linear range is not a problem and the lower propagation noise of the non-modulated PWFS can be fully exploited.
In contrast, the modulated PWFS is preferable for bright objects observed under poor seeing conditions, that is, when the extended linear range of this configuration is most critical.
Under all other conditions, performance is similar.}
\textcolor{black}{Interestingly, non-modulated PWFS have a better correction on low-order modes and the difference from modulated PWFS is greater for good seeing values} (see Fig. \ref{fig:modal_plot_R12_bin1}). 
\begin{table}
\caption{H band (1650 nm) SR as a function of natural guide star R magnitude and seeing for a non-modulated and a modulated PWFS. We note that for a sub-aperture size of 0.2, 0.4, 0.6, and 0.8 m,  the 500, 209, 90, and 54 modes are corrected respectively. \textcolor{black}{Refers to the 8m-class telescope case.}}
\label{Tab:results}
\begin{center}
\begin{small}
        \begin{tabular}{lccccc}
                \hline
                \textbf{seeing} & \textbf{R} & \textbf{sub-ap.} & \textbf{freq.} & \textbf{SR(H)} & \textbf{SR(H)} \\
                \textbf{[arcsec]} & \textbf{magn.} & \textbf{size [m]} & \textbf{[Hz]} & \textbf{no mod.} & \textbf{$r_{\mathrm{mod}}$=3$\lambda$/D} \\
                \hline
        0.4 & 8.0 & 0.2 & 1700 & 0.968 & 0.968 \\
        0.8 & 8.0 & 0.2 & 1700 & 0.899 & 0.902 \\
        1.2 & 8.0 & 0.2 & 1700 & 0.781 & 0.805 \\
                \hline
                
        0.4 & 12.0 & 0.2 & 900 & 0.957 & 0.949 \\
        0.8 & 12.0 & 0.2 & 900 & 0.870 & 0.863 \\
        1.2 & 12.0 & 0.2 & 900 & 0.731 & 0.735 \\
                \hline
                
        0.4 & 12.0 & 0.4 & 1200 & 0.923 & 0.917 \\
        0.8 & 12.0 & 0.4 & 1200 & 0.751 & 0.748 \\
        1.2 & 12.0 & 0.4 & 1200 & 0.518 & 0.514 \\
                \hline
                
        0.4 & 14.5 & 0.4 & 750 & 0.892 & 0.866 \\
        0.8 & 14.5 & 0.4 & 750 & 0.651 & 0.628 \\
        1.2 & 14.5 & 0.4 & 750 & 0.347 & 0.339 \\
                \hline
                
        0.4 & 14.5 & 0.6 & 1000 & 0.828 & 0.805 \\
        0.8 & 14.5 & 0.6 & 1000 & 0.474 & 0.458 \\
        1.2 & 14.5 & 0.6 & 1000 & 0.183 & 0.182 \\
                \hline
                
        0.4 & 16.5 & 0.6 & 250 & 0.737 & 0.679 \\
        0.8 & 16.5 & 0.6 & 250 & 0.288 & 0.279 \\
        1.2 & 16.5 & 0.6 & 250 & 0.066 & 0.065 \\
                \hline
                
        0.4 & 16.5 & 0.8 & 500 & 0.657 & 0.593 \\
        0.8 & 16.5 & 0.8 & 500 & 0.171 & 0.160 \\
        1.2 & 16.5 & 0.8 & 500 & 0.024 & 0.024 \\
                \hline
        \end{tabular}
\end{small}
\end{center}
\end{table}
%

\subsection{\textcolor{black}{Simulations with a compensation of the optical gain}}

\textcolor{black}{In this section, we considered the calibration error, namely}: the level of partial correction used for calibration is different with respect to the one found during operation.
\textcolor{black}{So, the optical gain is not correctly compensated by the SIMPC:} the calibration method we present in the previous sections \textcolor{black}{gives an accurate model of the PSWF response} when the value of seeing found during operation is the same value that was used for the calibration.

In principle, we could update the reconstruction matrix \textcolor{black}{regularly, following the level of seeing}, but here we want to explore the effect of an optical gain not equal to 1 and how it can be compensated for.
We considered a well established approach, tested on-sky on a 8m-class telescope: the one presented in \cite{2015aoel.confE..36E} and \cite{2020A&A...636A..88E} for the optical gain compensation, along with the one presented in \cite{2021MNRAS.508.1745A} for online gain optimization.
We note that other methods to deal with optical gain do exist, which may be considered \citep[see][]{2008ApOpt..47...79K,2018SPIE10703E..20D,2019A&A...629A.107D,2020A&A...644A...6C,Deo2021,Chambouleyron2021}, but \textcolor{black}{in this case,} we are not interested in finding a general solution for this issue. Our objective is to understand whether a non-modulated PWFS calibrated for a certain seeing value can operate with a different seeing value and if one method developed for a diffraction-limited IM can work also for the SIMPC presented in this paper.
We considered a sub-aperture size of 0.2 m, a bright NGS (R magnitude 8), seeing cases ranging from 0.4 to 1.4 arcsec, and interaction matrices with different levels of partial correction (computed for diffraction limited coniditions and seeing values between 0.4 and 1.4 arcsec).

\begin{figure}
    \begin{center}
    \begin{tabular}{c}
        \includegraphics[width=0.4\textwidth]{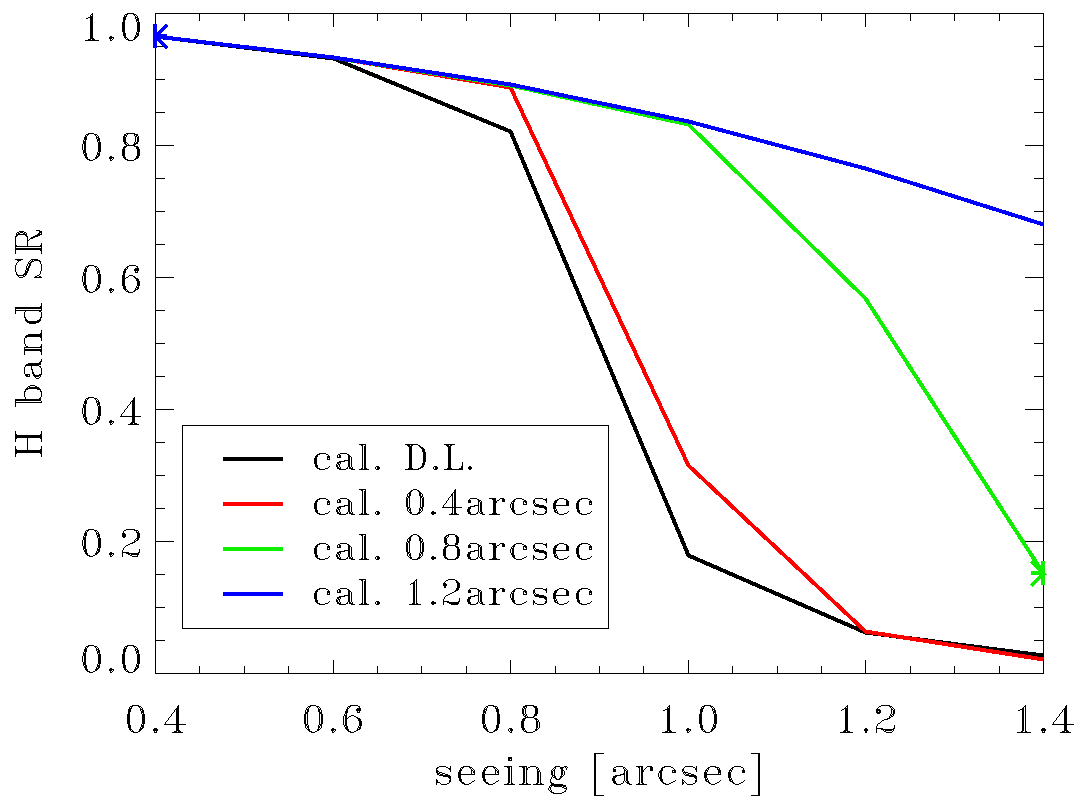}
    \end{tabular}
    \end{center}
         \caption{\label{fig:sr_seeing_gopt} H band (1650 nm) SR as a function of seeing for simulations with a non-modulated PWFS with bright NGS (R magnitude 8). Different colors correspond to different PWFS calibration: Black refers to a classic calibration in diffraction limited conditions, while the other colors are the SIMPC for different seeing levels. Please note that low values of SR correspond to unstable closed loops. Blue and green asterisks correspond to the cases illustrated in Figs. \ref{fig:res_seeing0.4_cal1.2} and \ref{fig:res_seeing1.4_cal0.8}. Refers
to the 8m-class telescope case.}
\end{figure}

\begin{figure}
    \begin{center}
    \begin{tabular}{c}
        \includegraphics[width=0.4\textwidth]{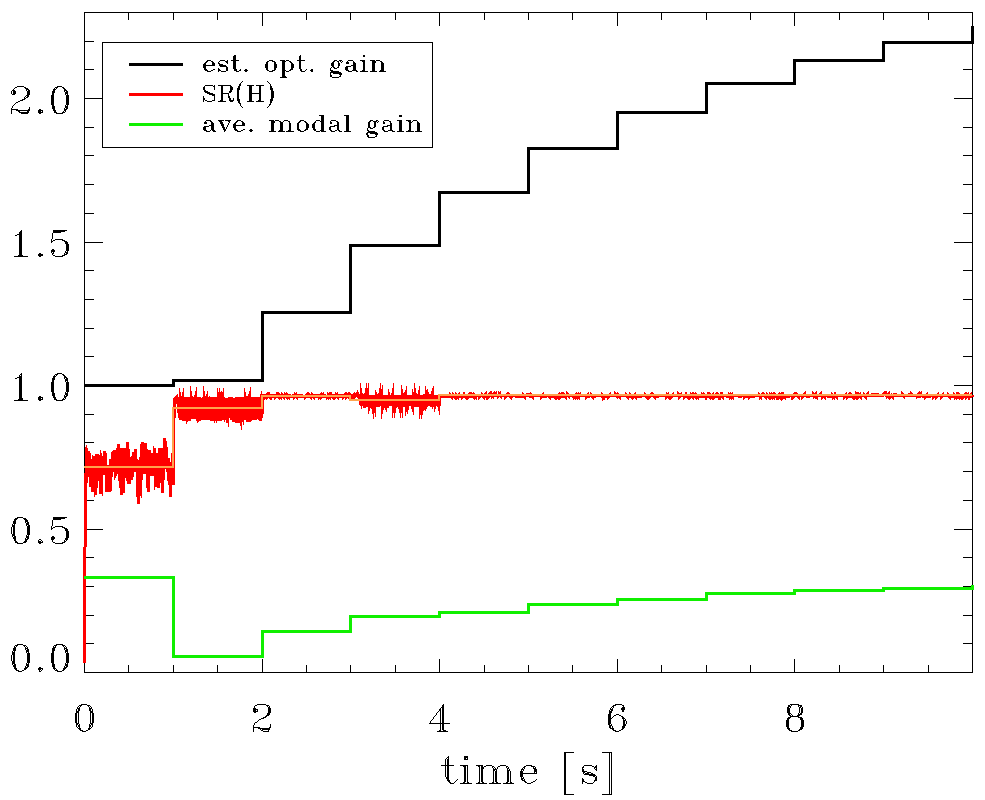}
    \end{tabular}
    \end{center}
         \caption{\label{fig:res_seeing0.4_cal1.2} Results for a simulation with a non-modulated PWFS with bright NGS (R magnitude 8) and with seeing of 0.4 arcsec with a calibration error: PWFS was calibrated for a seeing of 1.2 arcsec.
         The black line is the estimation of the optical gain (done on mode 30), red line is the H band (1650 nm) SR (orange line is the average on 1 s), and green line is the average value of the modal gain. \textcolor{black}{Refers
to the 8m-class telescope case.}}
\end{figure}

\begin{figure}
    \begin{center}
    \begin{tabular}{c}
        \includegraphics[width=0.4\textwidth]{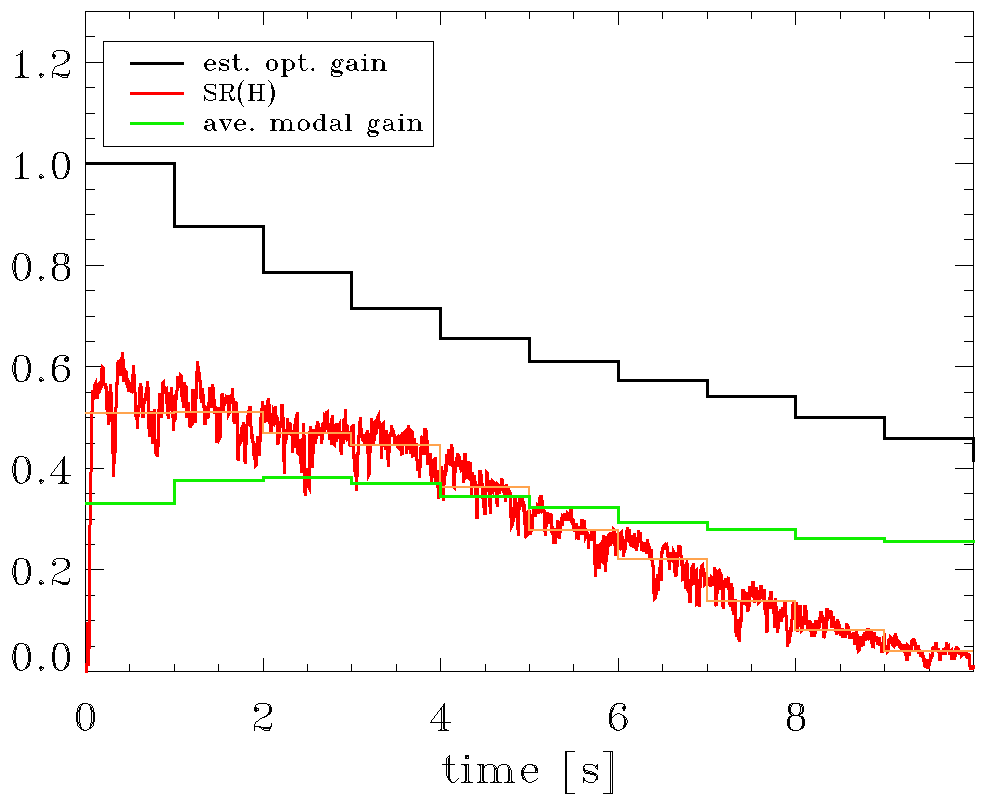}
    \end{tabular}
    \end{center}
         \caption{\label{fig:res_seeing1.4_cal0.8} Results for a simulation with a non-modulated PWFS with bright NGS (R magnitude 8) and with seeing of 1.4 arcsec with a calibration error: PWFS was calibrated for a seeing of 0.8 arcsec.
         The black line is the estimation of the optical gain (done on mode 30), red line is the H band (1650 nm) SR (orange line is the average on 1 s), and green line is the average value of the modal gain. \textcolor{black}{Refers
to the  8m-class telescope case.}}
\end{figure}

A summary of the results is presented in \textcolor{black}{Fig.} \ref{fig:sr_seeing_gopt}. Here, we can see that the interaction matrices computed for the worst seeing (performance given by SIMPC with seeing 1.2 arcsec is equivalent to the one with seeing 1.4 arcsec) are able to work with any seeing level with a performance comparable to the one given by the correct calibration, while the ones computed in diffraction-limited conditions or for good seeing become unstable for the case with the worst seeing.
A couple of \textcolor{black}{examples} of the results are shown in Figs. \ref{fig:res_seeing0.4_cal1.2} and \ref{fig:res_seeing1.4_cal0.8}. In the first case, the H band (1650 nm) SR reached is 0.966, equal to the one given by an IM without calibration error, while in the second case, the closed loop is not stable and the H band (1650 nm) SR goes to 0.
It is interesting to note that the estimated value of optical gain for the first case is above 1 because in this case the SIMPC has a lower sensitivity than the actual interaction matrix.

\textcolor{black}{As mentioned above, we note} that the result found here is not general, but it is valid for the optical gain compensation and modal gain optimization methods considered.
In particular, a different approach could allow for stable performance \textcolor{black}{even in the} case of a calibration \textcolor{black}{performed} with significantly lower levels of partial correction.
Nevertheless, this is \textcolor{black}{an interesting} result that \textcolor{black}{proves} that this issue can be solved and it could be a topic \textcolor{black}{ for further study in future work}.

\section{ELT case}\label{sec:elt}

In this section, we present an analysis carried out for the Extremely Large Telescope \citep[ELT,][]{2020SPIE11445E..1ET} PWFS system.
We considered the configuration reported in \cite{2018SPIE10703E..41X}: a system with 90$\times$90 sub-apertures (sub-aperture size 0.433 m) and a sensing wavelength of 798 nm in a telescope of 39m diameter without any spider shadow \textcolor{black}{(more details can be found in Table \ref{Tab:params} and in Appendix \ref{sec:simParams})}.

\begin{figure}
    \begin{center}
    \begin{tabular}{c}
        \includegraphics[width=0.4\textwidth]{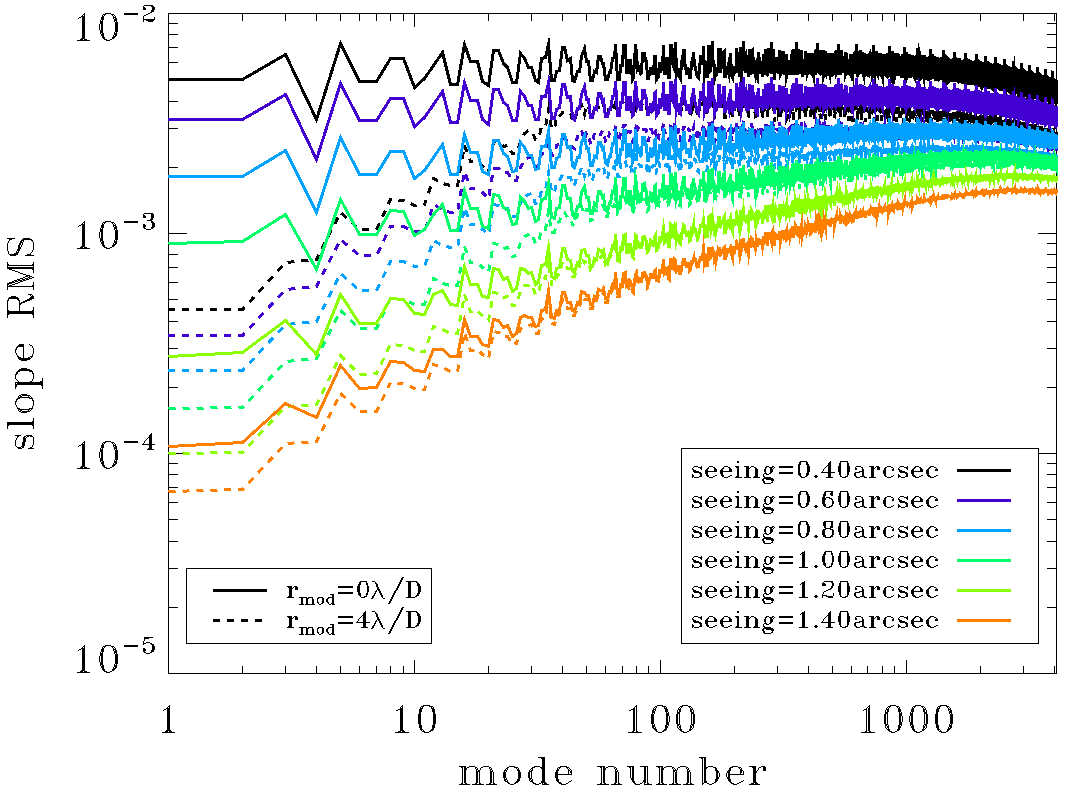}
    \end{tabular}
    \end{center}
         \caption{\label{fig:sens_b1_ELT} ELT PWFS sensitivity in partial correction as a function of mode number and seeing value expressed as the RMS of the slope for an aberration with \textcolor{black}{a wavefront} RMS of 1nm. Mode 1 is tip, 2 is tilt, and the other modes are Karhunen-Lo\`eve modes. Solid lines correspond to a non-modulated PWFS and dashed lines correspond to a PWFS with a modulation radius of 4$\lambda/D$. Sub-aperture size of 0.433 m. \textcolor{black}{Refers to the ELT case.}}
\end{figure}

\begin{figure}
    \begin{center}
    \begin{tabular}{c}
        \includegraphics[width=0.4\textwidth]{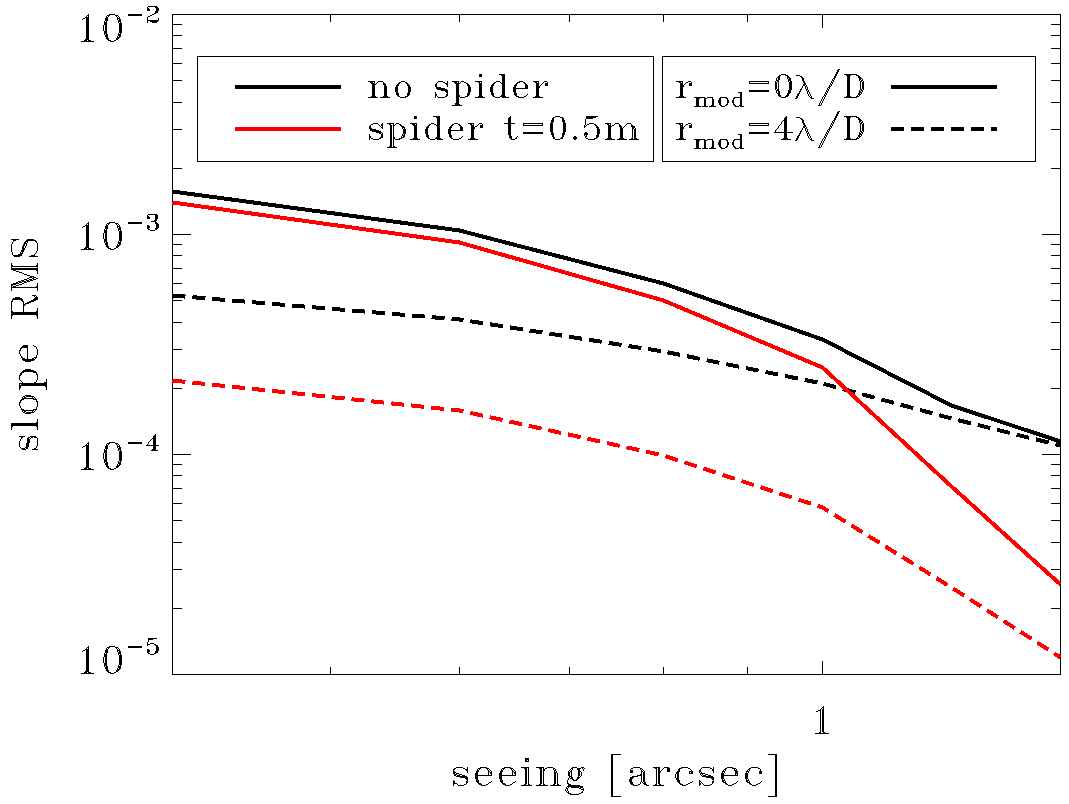}
    \end{tabular}
    \end{center}
         \caption{\label{fig:sens_pist_ELT} ELT PWFS sensitivity to a differential piston mode in partial correction as a function of seeing value expressed as the RMS of the slope for an aberration with \textcolor{black}{a wavefront} RMS of 1nm. Differential piston mode is the one produced by one petal of a 6 petals configuration with no gap and with a 0.5m gap. Solid line correspond to a non-modulated PWFS and dashed line correspond to a PWFS with a modulation radius of 4$\lambda/D$. Sub-aperture size is 0.433 m and seeing range is [0.4,1.4] arcsec. \textcolor{black}{Refers to the ELT case.}}
\end{figure}

We studied the sensitivity for this configuration as we describe in Sect. \ref{sec:sens}, with the results shown in Fig. \ref{fig:sens_b1_ELT}.
These results are similar to those of the 8m-class telescope case for a sub-aperture size of 0.4 m; between unmodulated and modulated PWFS, there is a factor of  9 for good seeing (1.5 for bad seeing) on the sensitivity of tip and tilt and a factor of 1.3 for good seeing (1 for bad seeing) on the sensitivity of the last mode.
We also replicated the sensitivity analysis for differential piston modes considering both a gap-less configuration and a configuration with 0.5 m \textcolor{black}{thick spider arms}.
\textcolor{black}{We can see in Fig. \ref{fig:sens_pist_ELT} that the non-modulated PWFS has an advantage over modulated PWFS and that this advantage is greater in the presence of spiders: a factor of 10 for good seeing values and a factor of 2 for bad seeing values when 0.5 m thick spider arms are present.}

Finally, we studied the performance of such system with a bright star (magnitude R=8) and we see that the results \textcolor{black}{exhibit consistent behaviors between the cases of the 8m- (reported in Sect. \ref{sec:results}) and the 39m-class telescope}: K band (2200 nm) SR for a seeing value of 0.4 arcsec is the same for non-modulated and modulated PWFS, 0.963, and there is an advantage of 0.03 K band (2200 nm) SR for modulated PWFS for a seeing value of 1.2 arcsec (0.730 w.r.t. 0.703).
\textcolor{black}{We note that these are preliminary results and this topic should be further studied to have a full assessment of the performance in the presence of the segmented pupil and the spider shadow of the ELT.}

\section{Conclusions}\label{sec:conclusion}

In this work, we focus on the non-modulated PWFS and demonstrate the way it can be used to effectively sense and correct atmospheric turbulence.
Our goal is to show why, how, and when the PWFS ought to be used with non-modulation.

The "why" is related to the \textcolor{black}{higher} sensitivity of the non-modulated \textcolor{black}{compared} to the modulated PWFS, \textcolor{black}{particularly} for differential segment piston sensing.
We \textcolor{black}{show} that the gain in sensitivity is greater for the best partial correction conditions (up to a factor of 10 for low order modes) -- thus, it is generally appropriate for good seeing and for AO systems with high sensing and correction sampling (\textcolor{black}{i.e.,} for extreme AO systems).

The "how" is based on a new approach to calibration of this sensor to model the PWFS response in the presence of a partial correction.
We call the interaction matrix computed with this approach the synthetic interaction matrix model with partial correction (SIMPC).
It is synthetic because the calibration is done in the simulations and the partial correction allows \textcolor{black}{to reduce} the model error between the PWFS response in calibration and in operation.
We show that this can \textcolor{black}{also be used} for the modulated case (\textcolor{black}{demonstrated with simulations) that it is effective under} a wide range of conditions for SCAO systems working \textcolor{black}{with} 8m- and 39m-class telescopes. We also present a possible approach to deal with a partial correction \textcolor{black}{that is different than the one} used in the calibration.

Finally, with respect to the "when," we \textcolor{black}{have shown} that the performance of a non-modulated PWFS is \textcolor{black}{very close to the one} of a modulated PWFS. \textcolor{black}{The non-modulated PWFS has a small advantage under conditions of good partial correction (high sensing and correction sampling and/or good seeing) and when the flux is low, while it has a small disadvantage under conditions of bad partial correction (low sensing and correction sampling and/or bad seeing) and when the flux is high.
Thus, in principle, the answer is that it can be always used, but the regime in which non-modulated PWFS performs better than modulated PWFS is when the contributions of fitting and temporal errors are low relative to noise.}
Another interesting case is the system with mirror segmentation: we did not study these systems in detail because this is not the focus of this work, but we showed that the gain in sensitivity in differential piston sensing can be \textcolor{black}{important} and is not associated with \textcolor{black}{a loss of} linearity.
We would like to point out an interesting fact  \textcolor{black}{\citep[at the heart of][]{2022JATIS...8b1502E}}, namely, that a modulated PWFS system can easily switch to a non-modulated case by simply turning off the tip-tilt modulator, so this  configuration can be used only when required.

A new calibration option for PWFS base AO system is available, which removes the constraints on the minimum value of modulation amplitude, potentially boosting the sensitivity and allowing for a better optimization of such a sensor.

\section*{Acknowledgements}
The authors thank C\'edric Plantet, Thierry Fusco, Benoit Neichel, Jean-Fran\c{c}ois Sauvage, Nicolas Levraud, Miska Le Louarn, Chistophe V\'erinaud, and Jean-Pierre V\'eran for fruitful discussions. 

\begin{appendix}
\section{\textcolor{black}{Details on the simulations}}\label{sec:simParams}

\textcolor{black}{In this section, we report several key details of the simulations presented in this work.
    We note that the 8m-class telescope case is equivalent to that of  single conjugate adaptive optics upgrade for LBT \citep[SOUL,][]{Pinna2019} and the ELT case is equivalent to the configuration reported in \cite{2018SPIE10703E..41X}.
A few notes to the parameters summarized in Table \ref{Tab:params} are:}
\begin{itemize}
    \item \textcolor{black}{Zenith angle is used to scale the distance of the layers from the entrance pupil. Seeing values reported in the text are always line-of-sight ones.
    \item Simulations are monochromatic.
    \item In the 8m-class telescope case different sub-aperture numbers are obtained by binning the detector pixels.    
    The maximum value is 40$\times$40 (sub-ap. size of 0.2 m), while 20$\times$20 (sub-ap. size of 0.4 m), 13$\times$13 (sub-ap. size of 0.6 m), and 10$\times$10 (sub-ap. size of 0.8 m) cases are obtained by binning the detector pixels by 2$\times$2, 3$\times$3, and 4$\times$4, respectively.
    \item The selected sub-pupil separation values derive from the glass pyramid geometry of SOUL. In SOUL there is a ratio between pixel on the diameter of the sub-pupil and separation that is 48/40.
    \item The influence functions for the 8m-class telescope case are measurements from the LBT adaptive secondary mirror \citep{2008SPIE.7015E..12R}. Instead, an ideal modal basis \citep[pure Zernike and Karhunen-Lo{\'e}ve modes][]{Wang:78} was used for the ELT case.
    \item Leaky integrators: we consider the discrete time case, $H(z)=g/(1-fz^{-1})$ \citep[see][]{Agapito2019}, their gains, $g$, are optimized \citep[modal gain optimization as described in][]{2021MNRAS.508.1745A}, and the forgetting factors, $f$, ranging from 1 to 0.9 \citep[as reported in section 7.2 of][noting that for the ELT case, $x$ is scaled by 0.15]{Agapito2019}.
    \item The delay in the case of the 8m-class telescope is between 2 (1700 Hz framerate) and 5 ms (250 Hz framerate), as shown in Figure 3 of \cite{2021MNRAS.508.1745A}.}
\end{itemize}

\textcolor{black}{Additional information on the simulations is provided below.}
\begin{itemize}
    \item \textcolor{black}{Turbulence is generated by moving phase screens.
    They have the same sampling as the pupil (8/220 m/pixel for the 8m-class telescope case and 39/512 m/pixel for the ELT case) and they are made of 32768$\times$32768 pixels.
    The phase screens are generated by computing the inverse fast Fourier transform of a complex array having amplitude equal to the square root of the von Karman power spectrum and having a spatially uncorrelated and uniformly distributed phase.
    Their amplitude is scaled considering line-of-sight seeing value and $C_n^2$ fraction.
    \item A geometrical propagation of the electric field is used from the turbulence layer to the telescope pupil. This propagation consider the geometrical information as the guide star altitude and position, the phase screens altitude, the deformable mirror (DM) conjugation altitude, the entrance pupil position (its conjugation altitude is 0 m), and the zenith angle, etc.
    \item The PWFS is implemented as described in \cite{PinnaTesi} with a sampling of the electric field in the focal plane of 1/3 $\lambda/D$ (where $\lambda$ is the sensing wavelength and $D$ is the telescope diameter).
    In the simulation presented in this work, we considered a monochromatic sensor.
    \item The electro-multipling process of the detector is simulated following \cite{2010ApOpt..49G.167C}.
    \item X- and y-sensor signals from the PWFS, also referred to in the text as slopes or vector of slopes, are calculated using a quad-cell-like calculation and normalized with the average intensity over all valid sub-apertures.
    So, for the $j$-th sub-aperture, the X-sensor signal is computed as:
    \begin{equation}
        s^{x}(j) = \frac{I_1(j) + I_4(j) - I_2(j) - I_3(j)}{\frac{1}{N_{sa}}\sum_{i=1}^4 \sum_{k=1}^{N_{sa}} I_i(k)} \; ,
    \end{equation}
    where $I_i$ is the intensity on the $i$-th sub-pupil, $k$ is the valid sub-aperture index, and $N_{sa}$ is the number of valid sub-aperture.
    \item The DM surface is computed as a linear combination of the influence functions (the conversion between surface and wavefront is given by a factor 2) projected on the pupil.
    \item In the cases reported in this work, the simulation time step is equal to the detector integration time.
    Non-integer delays are approximated with linear interpolation of the DM commands.
    \item Calibration amplitude is 20nm RMS for tip and tilt and then it is scaled by $1/\sqrt{N_{rad}}$, where $N_{rad}$ is the radial order of the considered mode.}
\end{itemize}

\end{appendix}

\bibliographystyle{aa}
\bibliography{biblio} 

\end{document}